\documentclass[useAMS,usenatbib]{mn2e}

\usepackage{graphicx}
\usepackage{url}
\usepackage{bm}
\usepackage{float}
\usepackage{placeins}
\usepackage{setspace}
\usepackage[fleqn]{amsmath}
\usepackage{longtable}
\usepackage{aas_macros}
\usepackage{amssymb}

\usepackage[utf8]{inputenc}
\usepackage[T1]{fontenc}
\usepackage{lmodern} 

\newcommand{\herschel}{{\it Herschel}}
\newcommand{\planck}{{\it Planck}}

\newcommand{\oracdr}{{\sc orac-dr}}
\newcommand{\smurf}{{\sc smurf}}

\title{SCUBA-2 follow-up of \textit{Herschel}-SPIRE observed \textit{Planck} overdensities}

\author[T. Mackenzie et al.]{Todd P. MacKenzie$^{1}$,
Douglas Scott$^{1}$,
Matteo Bianconi$^{2}$,
David L. Clements$^{3}$,\newauthor
Herve A. Dole$^{4}$, 
I. Flores-Cacho$^{5,6}$,
David Guery$^{4}$,
R. Kneissl$^{7,8}$,
G. Lagache$^{9}$,\newauthor
Francine R. Marleau$^{2}$,
L. Montier$^{5,6}$, 
N.~P.~H. Nesvadba$^{4}$,
Etienne Pointecouteau$^{5,6}$,\newauthor
G. Soucail$^{5,6}$
\\
$^{1}$Department of Physics \& Astronomy, University of British Columbia, 6224 Agricultural Road, Vancouver, BC V6T 1Z1, Canada \\
$^{2}$Institute of Astro and Particle Physics, University of Innsbruck, 6020 Innsbruck, Austria \\
$^{3}$Physics Department, Blackett Lab, Prince Consort Road, London SW7 2AZ, UK \\
$^{4}$Institut d'Astrophysique Spatiale, Univ. Paris-Sud \& CNRS, Univ. Paris-Saclay - IAS, b\^atiment 121, univ Paris-Sud, 91405\\ \,\,Orsay, France \\
$^{5}$Universit\'e de Toulouse, UPS-OMP, IRAP, 31028, Toulouse Cedex 4, France \\
$^6$CNRS, IRAP, 9 Av. Colonel Roche, BP 44346, 31028, Toulouse Cedex 4, France \\
$^{7}$Atacama Large Millimeter/submillimeter Array, ALMA Santiago Central Offices, Alonso de Cordova 3107, Vitacura,\\ \,\,Casilla 763 0355, Santiago, Chile\\
$^{8}$European Southern Observatory, ESO Vitacura, Alonso de Cordova 3107, Vitacura, Casilla 19001, Santiago, Chile\\
$^{9}$Aix Marseille Universit\'e, CNRS, LAM (Laboratoire d’Astrophysique de Marseille) UMR 7326, 13388, Marseille, France
}

\begin{document}
\maketitle

\begin{abstract}

We present SCUBA-2 follow-up of 61 candidate high-redshift \planck{} sources.  Of these, 10 are confirmed strong gravitational lenses and comprise some of the brightest such submm sources on the observed sky, while 51 are candidate proto-cluster fields undergoing massive starburst events.  With the accompanying \herschel{}-SPIRE observations and assuming an empirical dust temperature prior of $34^{+13}_{-9}$\,K, we provide photometric redshift and far-IR luminosity estimates for 172 SCUBA-2-selected sources within these \planck{} overdensity fields.  The redshift distribution of the sources peak between a redshift of 2 and 4, with one third of the sources having $S_{500}$/$S_{350} > 1$.  For the majority of the sources, we find far-IR luminosities of approximately $10^{13}\,\mathrm{L}_\odot$, corresponding to star-formation rates of around $1000$\,M$_\odot \mathrm{yr}^{-1}$.  For $S_{850}>8$\,mJy sources, we show that there is up to an order of magnitude increase in star-formation rate density and an increase in uncorrected number counts of $6$ for $S_{850}>8$\,mJy when compared to typical cosmological survey fields.  The sources detected with SCUBA-2 account for only approximately $5$ per cent of the \planck{} flux at 353\,GHz, and thus many more fainter sources are expected in these fields.

\end{abstract}


\section{Introduction}

Modern submm observatories, such as the JCMT \citep{holland}, BLAST\citep{pascale2008} and \herschel{} \citep{pilbratt2010}, have allowed us to view larger and larger portions of the submm sky to greater and greater depths, continually improving the statistics on this relatively new population of sources.  Of particular interest is the role of such sources in the census of global star-formation rates (SFRs) and understanding the driving force behind their intense star-formation activity.  While some may be triggered by mergers \citep[e.g.][]{clements1996,murphy1996,sanders1996}, others may simply be at the bright end of what has been named the ``main sequence'' of galaxies \citep{noeske2007,daddi2007,elbaz2011}.  Most wide-field cosmology surveys try to characterise this population as a whole, but it is important to consider the effects of galaxy environment on star-formation.  Due to detection techniques, most known clusters are at redshifts below that of the peak of global star-formation, and the star-formation rate of cluster members has been quenched through various physical processes, although galaxies falling into their gravitational potential wells for the first time may still experience an increase in star-formation \citep{verdugo2008,braglia2009,braglia2011}.  The Sunyaev-Zeldovich effect, providing largely redshift-independent selection, has been used to detect hundreds of clusters out to $z\sim1.5$ \citep[e.g.][]{plancksz1,actsz,plancksz2,plancksz3,bleem2015}.  Similarily, the Spitzer
Infrared Array Camera (IRAC) Shallow Survey used 4.5\,$\mu$m selected sources and photometric redshifts to identify over a hundred candidate clusters with redshifts ranging from 1 to 1.7 \citep{iracsurvey}.

A complementary high-$z$ cluster detection technique is to look for regions of exceptional star-formation.  Due to the density of such objects on the sky, large areas must be probed in order to find a significant sample, thus all-sky surveys are needed.  \planck{} (with its 5 arcminute beam that closely matches the expected size of a forming galaxy cluster at $z\sim2$--$4$), along with its all-sky coverage, is an excellent observatory for finding such objects.  Indeed, followup of objects already observed by \herschel{} from the \planck{} Early Release Compact Source Catalog \citep{ercsc} revealed four star-forming clusters out to a redshift of 2.3 \citep{clements2014}.  However, the area mapped by \herschel{} is only about 90\,deg$^2$, and thus there remains a much larger portion of the sky that is mapped by \planck{}, but not \herschel{}.  The search for further star-forming cluster candidates not yet observed by \herschel{} has already been performed and the first results can be found in \cite{planckselection}.  The catalogue contains 2151 objects comprising the \planck{} high-$z$ (PHZ) catalogue found using 26 per cent of the sky.  

Two methods were used to generate a list of potential high-$z$ targets to follow-up with \herschel{} (see \citealt{herve} 
for details).  The first of the two methods is explicitly detailed in \cite{planckselection}, leading to the PHZ list, which will be available on the Planck Legacy Archive.  This method uses Cosmic Microwave Background (CMB) and Galactic-cirrus-cleaned \planck{} maps at 353, 545 and 857\,GHz, using the 26 per cent of the sky that is the least contaminated by Galactic sources.  S/N $>$ 5 sources were identified in a 545\,GHz ``excess'' map, defined to be the 545\,GHz map with a linear interpolation between the 353 and 857\,GHz maps subtracted.  On top of this, S/N $>$ 3 detections were required at 353, 545 and 857\,GHz.  To remove cold Galactic cores and extragalactic radio sources, only detections with $S_{545}/S_{857}>0.5$ and $S_{353}/S_{545}<0.9$ were retained.  The second method used the Planck Catalogue of Compact Sources \citep[PCCS,][]{pccs} and a selection method based on the work of \cite{negrello}.  Here, 52 per cent of the sky was used, based on the 857\,GHz Galactic mask, and sources with S/N $>$ 4 at 545\,GHz were selected from the catalogue.  From this list, sources were only retained with $S_{857}/S_{545}<1.5$ and $S_{217}/S_{353}<1$, and which were not identified as a local galaxy, a bright radio source or Galactic cirrus in either the NASA/IPAC Extragalactic Database (NED), ALADIN, or {\it IRAS} maps.  The result is a list of high-$z$ candidate sources, selected to have apparent redshifted flux densities peaking between 353 and 857\,GHz.  Included in this is a combination of strongly lensed sources, proto-clusters undergoing massive starbursts, chance overdensities of star forming galaxies, and perhaps a few Galactic interlopers.  The fraction of objects in the various categories is currently unknown, which is why follow-up observations are critical.


A total of 228 of these candidates were observed by \herschel{} using the Spectral and Photometric Imaging Receiver \citep[SPIRE,][]{spire}.  This instrument, with a beamsize 16 times smaller than \planck{}'s, has the ability to resolve the \planck{} candidates into either single bright point sources or many fainter sources.  The former were shown by \cite{gems} to be among the brightest strongly lensed sources on the sky;  11 out of 15 of these bright sources were followed-up (two more were previously known, as discussed in \citealt{fu} and \citealt{combes}, while the remaining two are in the far south) with a host of instruments, including SCUBA-2 at 850\,$\mu$m, spectroscopic observations using the wide-band heterodyne receiver Eight MIxer Receiver (EMIR) at the Institut de Radioastronomie Millim\'etrique telescope (IRAM) and SMA 850\,$\mu$m interferometry, to confirm their lensed nature.  Their redshifts range from 2.2 to 3.6, with peak flux densities from 0.35 to 1.14\,Jy, and they have apparent far-IR luminosities up to $3\times10^{14}\,\mathrm{L}_\odot$.  Due to their extra-ordinary flux densities and far-IR luminosities, these sources have been named ``\planck{}'s dusty GEMS'' (Gravitationally Enhanced subMillimetre Sources). 

The first results covering the other class of source, namely the ``overdensity fields,'' are presented in \cite{herve}.  Significant enhancements in the surface density of sources were found at 350 and 500\,$\mu$m, with the majority of sources peaking at 350\,$\mu$m.  Assuming an average dust temperature of 35\,K, \cite{herve} found a typical redshift of 2 for the overdensity fields, with average far-IR luminosities of around $4\times10^{12}\,\mathrm{L}_\odot$ per SPIRE source.  These \planck{}-selected sources may be high redshift proto-clusters undergoing rapid starformation, although some may also be chance line-of-sight alignments.  Without spectroscopic redshift estimates of the objects within these overdensities, it is impossible to distinguish between these two possibilities.

The analysis here focuses on the SCUBA-2 observations, based on 61 of the 228 \herschel{} fields, which have been followed up with 850\,$\mu$m observations at the JCMT.  10 of these fields are observations of \planck{}'s dusty GEMS and are detailed in \cite{gems}.  The 51 ``overdensity fields'' are discussed here.  The more favourable ``k-correction'' \citep{franceschini,blain} at 850\,$\mu$m means that we have a significantly less biased view of the redshift distribution of the overdensity fields than \herschel{}-SPIRE, and a greater sensitivity to sources at redshifts $\gtrsim3$.  We use the {\sc SEDeblend} method adapted from \cite{Mackenzie2014}, as described in more detail in \cite{alesspaper}, to fit modified blackbody SEDs to the SCUBA-2-detected sources.  This method deblends SEDs directly from the maps, as opposed to a traditional approach where flux densities are deblended first, followed by subsequent SED fitting to the deblended fluxes.  Degeneracy information between neighbouring confused sources is therefore retained, which is reflected in the fit parameters constraints, and more robustly deals with the confused nature of the \herschel{}-SPIRE imaging. To do this, we use the SCUBA-2 positions and fluxes, as well as the \herschel{}-SPIRE imaging.  We use a relatively weak prior on dust temperature to break its degeneracy with redshift, giving us useful constraints on both redshift and far-IR luminosity. Throughout we employ a $\Lambda$CDM cosmology with $\Omega_\Lambda=0.692$, $\Omega_\mathrm{m}=0.308$, and $H_0=67.8\,\mathrm{km}\,\mathrm{s}^{-1}\,\mathrm{Mpc}^{-1}$ \citep{planckcosmology}.

\section{The Planck candidates follow-up}

\subsection{SCUBA-2 follow-up}

\begin{figure*}
\begin{center}
\includegraphics[width=12cm]{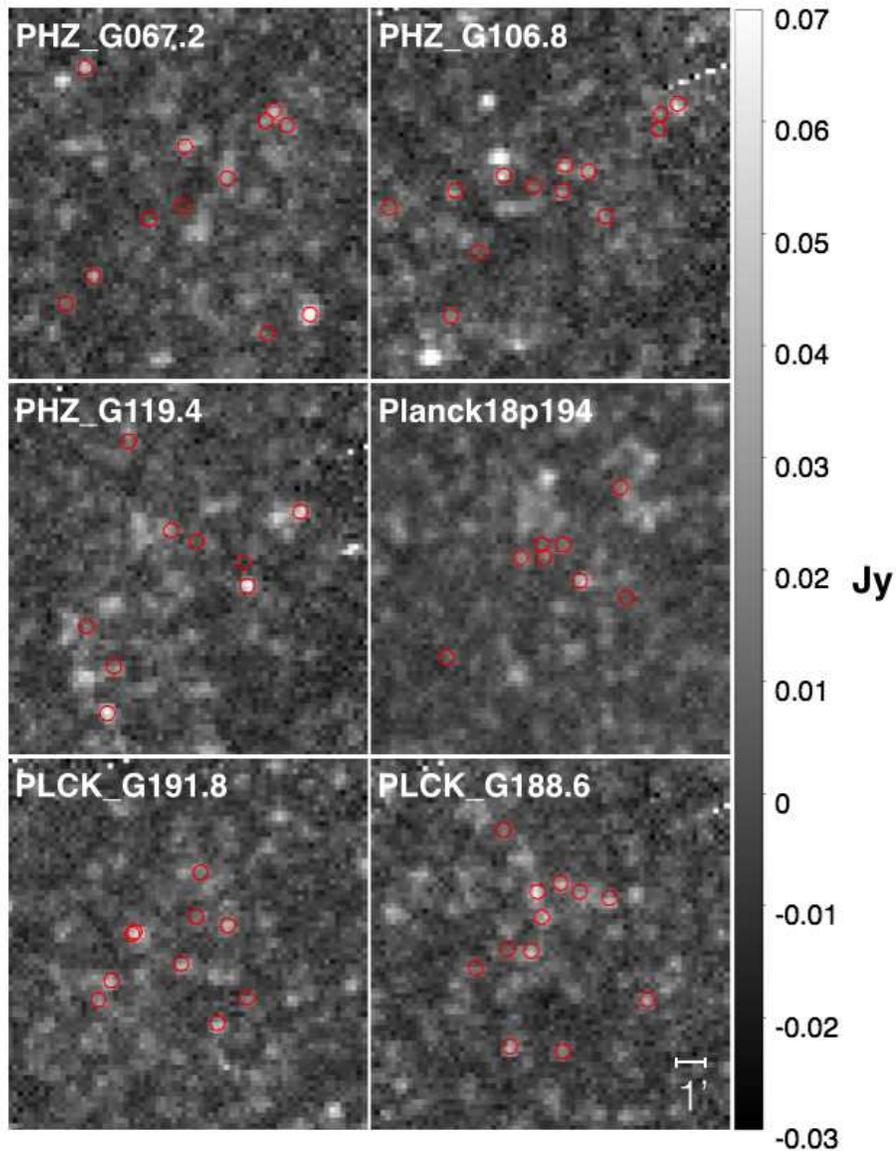}
\end{center}
\caption{SCUBA-2 850\,$\mu$m positions (red circles) plotted on the 350\,$\mu$m {\it Herschel} SPIRE images for the six fields with the densest concentration of SCUBA-2 sources. For the majority of the sources, we find far-IR luminosities of approximately $10^{13}\,\mathrm{L}_\odot$, corresponding to star-formation rates of around $1000$\,M$_\odot \mathrm{yr}^{-1}$.  For $S_{850}>8$\,mJy sources, we show that there is up to an order of magnitude increase in star-formation rate density and an increase in uncorrected number counts of approximately $6$ for S$_{850}>8$\,mJy when compared to typical cosmological survey fields.}
\label{examplefield}
\end{figure*}

51 overdensity fields observed with \herschel{}-SPIRE have been followed-up with SCUBA-2 850\,$\mu$m observations at the JCMT (project codes M12AC19, M13AC22, M13BC05, M13BU09 and M14AC02) with approximately 10 arcminute diameter ``daisy'' scans.  The initial observation strategy involved 20 minute scans repeated twice in grade 3 weather, while later observations used 30 minute scans repeated three times in grade 2 and 3 weather.  The observations were reduced using \smurf{}\ \citep{smurf} called from the \oracdr\ pipeline \citep{oracdr} using the standard blank-field map-making recipe optimised for finding point-sources.  Readings from the JCMT water-vapour monitor (WVM, \citealt{wvm}) and the scaling relations found by \cite{dempsey2010} were used to correct for atmospheric extinction.  

To facilitate finding and extracting point-sources, we use the standard ``matched-filter'' provided by \oracdr{}.  This procedure subtracts a 30 arcsecond smoothed map from a map convolved with the point spread function of the telescope and applying a correction factor such that point sources return the expected flux density.  This correction factor accounts for source attenuation by both matched-filtering and bolometer time-series high-pass filtering by \smurf{}, and is estimated by injecting simulated sources into bolometer time-series and recovering their flux density after map-making, matched-filtering and source extraction.  The minimum rms depths of each field range from 1.5 to 3\,mJy (instrumental noise), with a median of 1.9\,mJy for point sources in the matched-filtered images.  We extract peak flux densities and positions from these maps to generate a catalogue of 172 SCUBA-2-detected sources with S/N\,$>$\,4 in 1.20\,deg$^2$ of sky for the 51 \planck{} overdensity fields.  Fig.~\ref{examplefield} shows SCUBA-2 positions plotted over the 350\,$\mu$m Herschel SPIRE images for six fields with the densest concentration of SCUBA-2 sources.  We also require a flux density uncertainty of less than 4\,mJy for every source, since higher noise regions near the edges of the maps are more likely to contain artefacts of the map-making procedure.  Of the 1.2\,deg$^2$, 0.69\,deg$^2$ was within the \planck{} beams, which we define to be the area in the \planck{} 353\,GHz map with flux density greater than half the peak flux density of the \planck{} source, as in \cite{herve}.  Of the 172 SCUBA-2-detected sources, 138 are located within the \planck{} beam.  Table~\ref{sourcelist} list the source positions and flux densities, as well as constraints on their far-IR luminosities and redshifts.  We refrain from deboosting the flux densities of our catalogue since the number counts in these regions will differ from those in cosmological fields, and hence it is hard to estimate the effects of confusion \cite[see e.g.][]{coppin2006,scott2010}.

In order to asses the number of spurious sources within our catalogue, we perform the same source extraction procedure described above to search for negative sources within our maps.  For this process, we avoid negative sources associated with ``negative bowls'' surrounding bright positive sources caused by the matched-filter.  We find a total of 28 negative sources that satisfy our search criteria, which is higher than the roughly 2 expected spurious sources given the area observed and chosen signal-to-noise threshold.  However, if we consider the effects of confusion noise of about 0.7\,mJy per beam, the number of spurious negative sources with S/N\,>\,4 is reduced to 15.  We believe this excess of significantly negative sources is the result of relatively few repeats of the observation fields and resulting map-making artefacts.  Because the probability of detecting both false negatives and false positives should be equal, we expect an equal number of false positives in our catalogue.  By characterising the properties of the false negative sources, we can correct for the false positive contributions when estimating the redshift distribution and global co-moving star-formation density of the sources in our catalogue.  More information about these false negatives and their properties is given below.

\subsection{\textit{Herschel}-SPIRE data}

All our observed fields have accompanying \herschel{}-SPIRE observations at 250, 350 and 500\,$\mu$m.  These observations have been reduced using {\sc hipe} 10 \citep{hipe}, with the details given in \cite{herve}.  The images have instrumental noise levels of 7.7, 6.3 and 7.6\,mJy per pixel using the standard pixel sizes of 6, 10 and 14 arcseconds at 250, 350 and 500\,$\mu$m, respectively.  Thus, the noise levels in the images are near the confusion limit of \herschel{}-SPIRE \citep{confusion}, which we estimate to be 7.4, 8.7 and 8.6 at 250, 350 and 500\,$\mu$m, respectively for our fields.  Confusion noise is the result having an average of more than one source per beam, creating an undulating background.  This limits the precision to which you can know the flux density of a source in the observations and the quoted values are one sigma uncertainties.

\section{SED model and fitting}
\label{model}

We use the {\sc SEDeblend} method as described in \cite{alesspaper} to constrain SED parameters for each source.  This method forward-models every source SED into the image plane to reconstruct simulated observations at each wavelength simultaneously.  This model is then fit using Monte Carlo Markov chain (MCMC) Metropolis-Hastings algorithm \citep{mcmc1,mcmc2} and uses a likelihood calculation that accounts for the effects of confusion noise in the \herschel{}-SPIRE observations.  The version of {\sc SEDeblend} applied here has a few key differences.  First, since we do not have a catalogue of nearby sources to use for deblending, we instead use blank sky data to estimate the sky covariance matrix.  Specifically, we turn to the GOODS-North HerMES field used in \cite{Mackenzie2014}, with Gaussian random noise added in quadrature to the instrumental noise, so that the images contain the same noise levels as the \planck{} overdensity fields.  While the confusion noise in our fields will be higher than that of a blank field, we estimate that the total per pixel noise in our fields is only 10 to 15\% higher than blank fields observed for the same amount of time.  We treat the SCUBA-2-detected sources in the same way as the ALMA-resolved LESS sources in \cite{alesspaper}, using the source positions and flux density estimates at 850\,$\mu$m (although source positions are not as well constrained, of course). To account for source position uncertainty, we allow source positions to vary, with a 3 arcsecond positional prior, applied to the radial offset, up to a maximum of 10 arcseconds.  Such positional errors are typical for $5\sigma$ SCUBA-2 850\,$\mu$m sources \citep[e.g.][]{simpson2015}.  In addition to allowing for source position uncertainties, we allow the telescope pointing to vary with a 1.5 arcsecond prior.  The former positional prior accounts for source position uncertainty due to instrumental noise and applies to sources individually, while the latter accounts for telescope pointing uncertainty and affects all sources within a field in the same manner.  An additional 5 per cent calibration uncertainty is applied to the SCUBA-2 flux estimates \citep{scuba2calibration}. Far-IR luminosities are calculated by integrating the model SED from 8 to 1000\,$\mu$m. 

While the ALMA LABOCA ECDFS Submm Survey (ALESS) sample had independent photometric redshift estimates \citep{simpson}, our catalogue does not.  Instead, we apply a prior on dust temperature in order to generate estimates for both redshifts and far-IR luminosities of our sample.  Of course, redshift estimates would change if we used a different dust temperature prior; however, we make sure to choose a prior distribution with a realistic width, and to the extent that the dust temperature does not change dramatically with redshift, our far-IR luminosity estimates should be reasonably accurate.  Using sources above the 4.2\,mJy flux density limit of the LABOCA ECDFS Submm Survey (LESS) for the ALESS follow-up in \cite{alesspaper}, we find a dust temperature distribution of $34^{+13}_{-9}$\,K (using the 16th, 50th and 84th per centile values), and we adopt this fairly broad distribution as our dust temperature prior.  Note that this represents the width of the distribution and not the error on the mean of the distribution.  This central dust temperature and range of this prior is consistent with previous estimates for sources selected at 850\,$\mu$m \citep[e.g.][]{chapman2005,swinbank}.

In addition to fitting SEDs to the SCUBA-2-selected sources in the \planck{} overdensity fields, we apply the same method to SCUBA-2-selected sources from the Cosmology Legacy Survey \citep[CLS,][]{simpson2015} within the Ultra Deep Survey (UDS) field \citep{uds}.  This field also has accompanying \herschel{}-SPIRE observations from HerMES \citep{hermespaper}.   By performing an identical analysis on the sources that are detected in this field, we are able to perform a direct comparison with the \planck{} overdensity fields.  In addition to the availability of both SCUBA-2 and \herschel{}-SPIRE observations, this field was chosen because the data are deeper and the area of the sky surveyed is almost identical to that covered by the \planck{} overdensity fields.  Before fitting SEDs to these sources, we add Gaussian random noise in quadrature to the instrumental noise of the HerMES SPIRE images to give it the same noise levels as the \planck{} overdensity fields, while accounting for the difference in pixel sizes.  This catalogue contains 619 SCUBA-2-detected sources within 1.05\,deg$^2$ of sky, with an average source flux density uncertainty of 1.2\,mJy.  The large number of sources found in this field is due to the longer integration time and steep submm number counts.  Similarily, we find 26 negative sources within the UDS field.

\section{SED fitting results}

\begin{figure*}
\begin{center}
\includegraphics[width=11cm,trim = 25mm 10mm 15mm 10mm, clip, angle=-90]{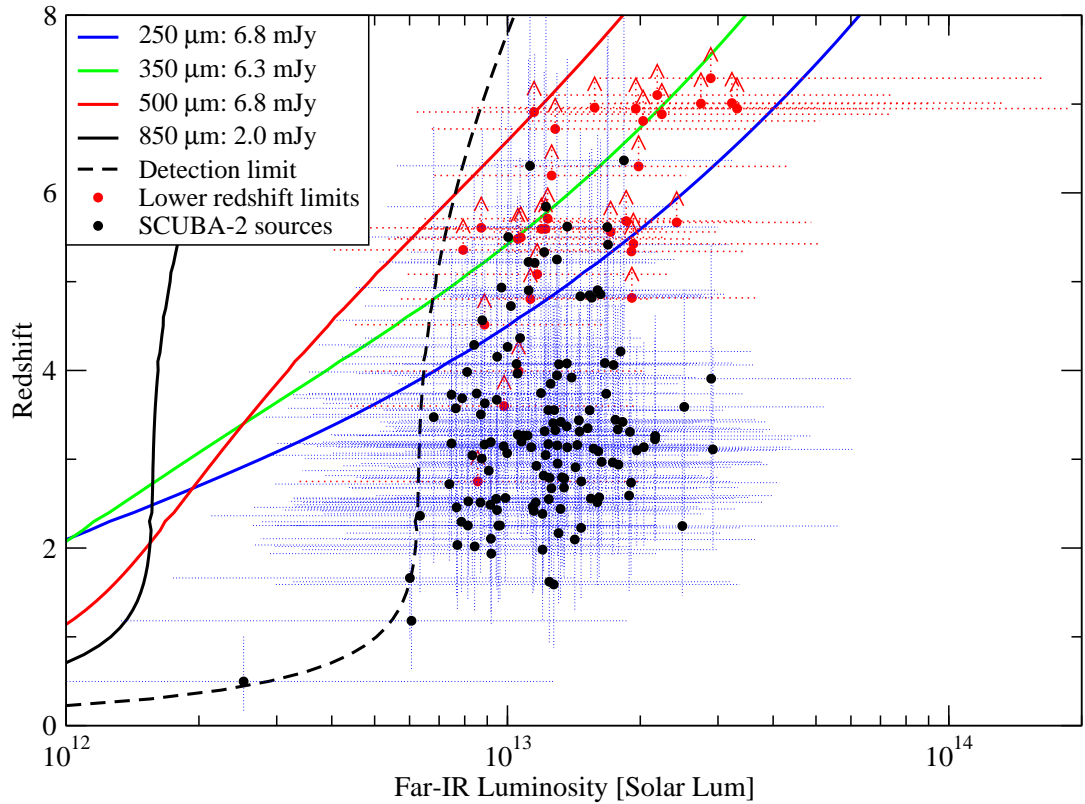}
\end{center}
\caption{Constraints on far-IR luminosities and photometric redshifts for the sample of SCUBA-2-detected sources within the \planck{} ``overdensity fields.''  We report the median values from the MCMC chains and plot 68 per cent confidence intervals for both far-IR luminosities and photometric redshifts.  Photometric redshift fits for sources with redshifts greater than 6.5 are not possible due to the peak of the SED being located outside the wavelength coverage; these sources are shown with red points representing the 16 per centile of the distributions as a 1$\sigma$ lower confidence limit.  Solid lines denote approximate 1$\sigma$ 2.0, 6.8, 6.3, and 5.8\,mJy limits at 850, 500, 350, and 250\,$\mu$m, respectively, for a 34\,K dust temperature and $\beta=1.5$ dust emissivity modified blackbody \protect\citep[the former is a representative 850\,$\mu$m point source flux density uncertainty and the later are the \herschel{}-SPIRE confusion limits from][]{confusion}.  Note that our source list is expected to have a rather large fraction of spurious sources (perhaps 15 per cent).   When fitted, many sources get constrained to redshifts greater than 6.5, since the SPIRE images contain no nearby counterparts.  For redshifts less than 6.5, we only expect nine spurious source, based on searching for negative peaks.}
\label{planckresults}
\end{figure*}

The results of the SED fitting are shown in Fig.~\ref{planckresults} and listed in Table~\ref{sourcelist} along with 68 per cent confidence intervals.  Due to the wavelength coverage of the data, we are not able to constrain redshifts for sources with redshifts greater than about 6.5, and for those sources we report the 16th per centile of the distributions as a 1$\sigma$ lower confidence limit.  For high redshift sources, these limits are affected by a hard prior that sources cannot have a redshift greater than 10.  Using our temperature prior of $34^{+13}_{-9}$\,K, we achieve a photometric redshift uncertainty of $\delta z / (1+z)\approx0.28$, with 68 per cent confidence intervals skewed to higher redshifts due to the asymmetric prior.  In addition to fitting SEDs to the 172 SCUBA-2-detected sources, we perform a further test of our method by fitting SEDs to the 28 negative sources in the map above the 4$\sigma$ cutoff (treating the negative flux densities at 850\,$\mu$m as positive).  Since there should be no \herschel{} counterparts, the majority of these sources are constrained to the high redshift region of Fig.~\ref{planckresults}, with 19 of the 28 negative sources falling into this category.  Of the 172 positive sources in our catalogue,  32 sources have median redshifts greater than 6.5, and should be considered suspect.  Only nine negative sources coincidentally have \herschel{} counterparts and redshift estimates lower than 6.5; this lack of spurious sources with low inferred reshifts suggests that sources with lower redshifts should be considered more reliable (approximately 6 per cent contamination).

\begin{figure*}
\begin{center}
\includegraphics[width=11cm,trim = 25mm 5mm 15mm 10mm, clip, angle=-90]{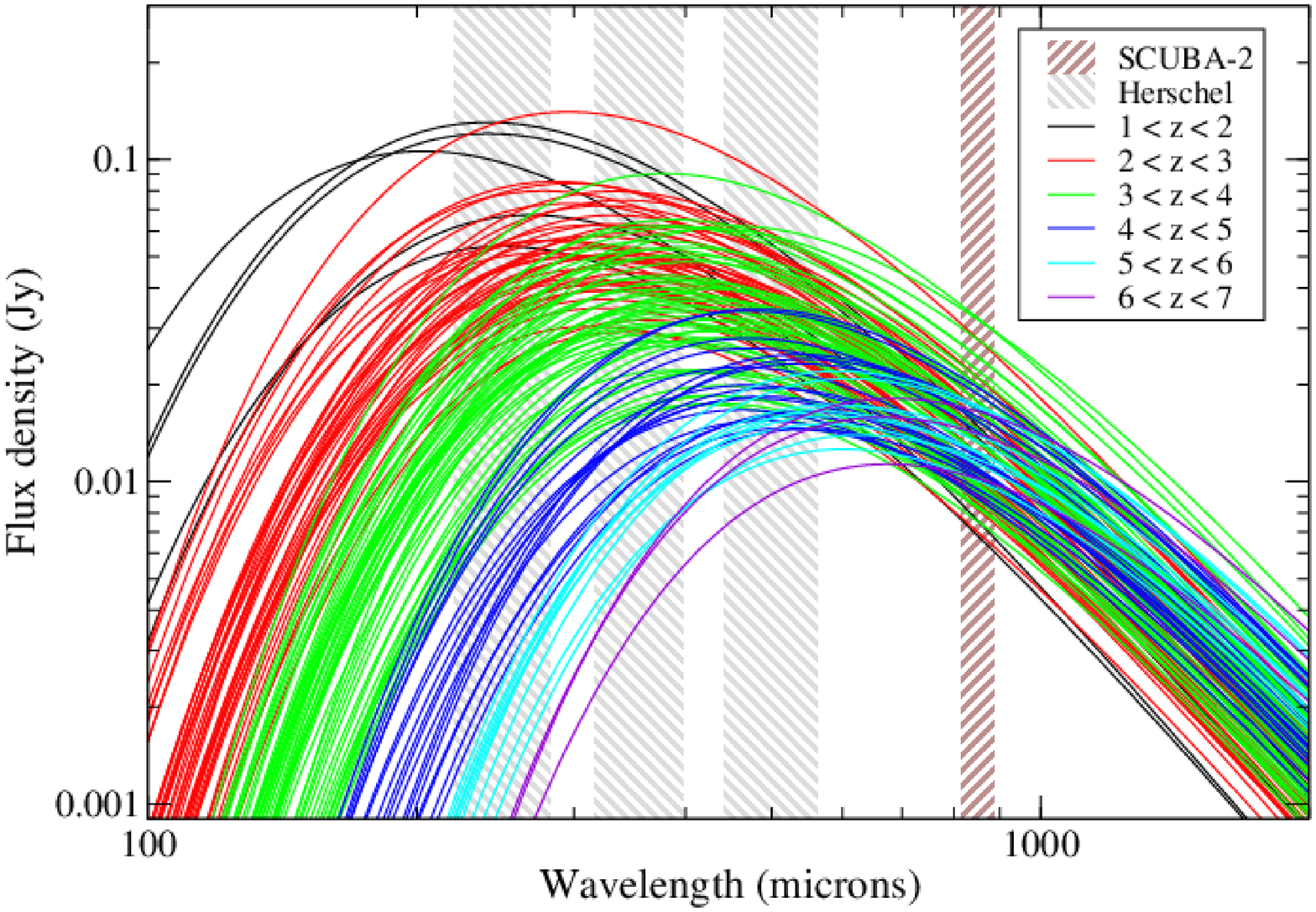}
\end{center}
\caption{Model SEDs of the best-fit results for SCUBA-2 sources with constrained redshifts greater than 1.  Grey and brown hatched areas denote approximate {\it Herschel}-SPIRE and SCUBA-2 band passes.  These bandpasses are ideally located for constraining model SED parameters for sources ranging from a redshift of 1 through 7, given a source dust temperature of around 34\,K. }
\label{sedexamples}
\end{figure*}

\section{Discussion}

In Fig.~\ref{planckresults}, the majority of well-constrained sources have far-IR luminosity estimates of around $ 10^{13}\mathrm{L}_\odot$, corresponding to SFRs of roughly $1000\,\mathrm{M}_\odot\,\mathrm{yr}^{-1}$.  On average, these sources are more luminous than those found in \cite{herve}, which have an average of $4\times10^{12}\mathrm{L}_\odot$ (assuming a dust temperature of 35\,K), although this is easily explained when considering the selection effects.  A representative 2\,mJy $1\sigma$ detection limit at 850\,$\mu$m for a 34\,K source is plotted in Fig.~\ref{planckresults} with a solid blue line, along with \herschel{}-SPIRE confusion limits for a source of the same dust temperature and dust emissivity.  From these detection limits, it is clear that SCUBA-2 is more sensitive to sources at $z\gtrsim$\,3.  While \cite{herve} found that only 3.5 per cent of \herschel{}-SPIRE 350\,$\mu$m-detected sources peak in the 500\,$\mu$m waveband, we find that 33 per cent of the SCUBA-2-detected sources have SED models with predicted 500\,$\mu$m to 350\,$\mu$m flux density ratios greater than 1.

\begin{figure*}
\begin{center}
\includegraphics[width=10cm,trim = 25mm 0mm 15mm 10mm, clip, angle=-90]{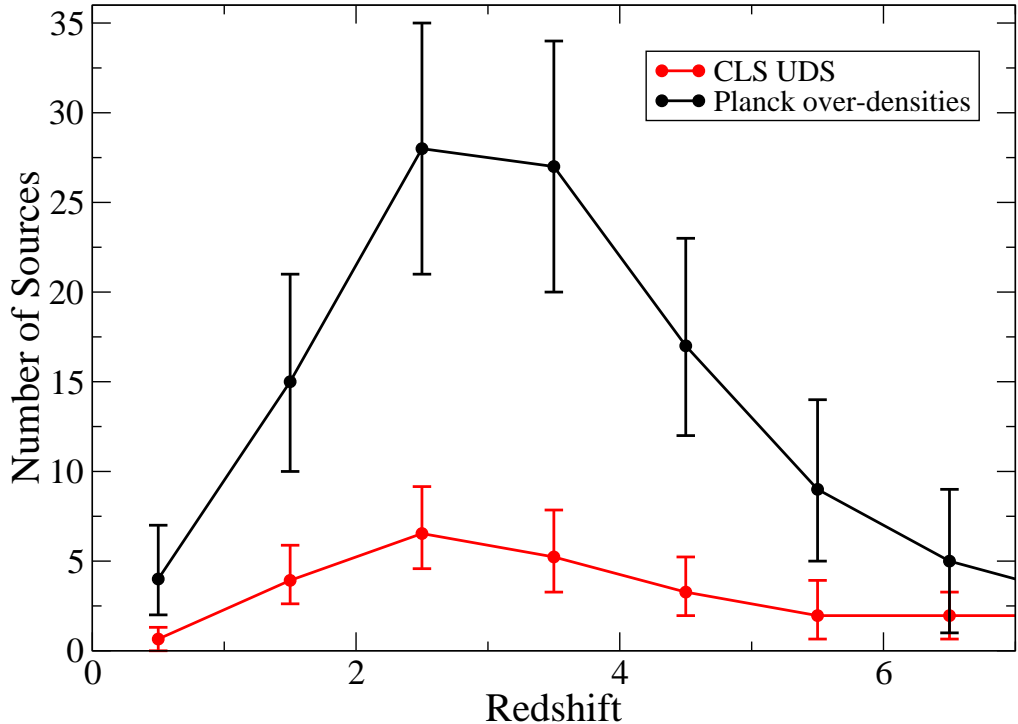}
\end{center}
\caption[Redshift distribution of SCUBA-2-detected sources, assuming a dust temperature prior of $34^{+13}_{-9}$\,K, for the \planck{} overdensity sample within the \planck{} beam.]{Redshift distribution of SCUBA-2-detected sources, assuming a dust temperature prior of $34^{+13}_{-9}$\,K, for the \planck{} overdensity sample within the \planck{} beam.  Also plotted is the expected distribution of CLS UDS sources given the same sky coverage and similar selection and flux-boosting effects.  Error bars are 68 per cent confidence intervals derived from bootstrapping the sample and have been corrected for estimated contributions from spurious sources.  The \planck{} overdensity fields contain a factor of about 4 more sources than the CLS UDS field, when given a similar selection function and sky coverage, but the shape of the distribution appears similar.}
\label{redshiftdistribution}
\end{figure*}

Fig.~\ref{redshiftdistribution} shows the estimated redshift distribution of the SCUBA-2 catalogues within the \planck{} beam for the \planck{} overdensities, with 68 per cent confidence intervals.  Also plotted is the expected CLS UDS redshift distribution, given the same survey area and selection function.  This plot is generated by bootstrapping the Monte Carlo Markov chains (MCMC) generated by the {\sc SEDeblend} method and is corrected for contributions from spurious sources by subtracting the redshift distribution of the negative SCUBA-2 sources.  The majority of these negative sources have estimated redshifts greater than 6 due to the absence of associated \herschel{}-SPIRE detections and their subtraction should correct the estimated redshift distribution for contribution from spurious (positive) SCUBA-2 detections.  

In order to give the CLS sources a similar selection function and flux density boosting as the overdensity fields, we add Gaussian random noise to the CLS SCUBA-2 fluxes, such that the distribution of flux density uncertainties matches the distribution of randomly selected points within the \planck{} overdensity fields with a flux density uncertainty $<4$\,mJy and that are within the \planck{} beam.  It is important to note that this redshift distribution estimate is actually a convolution between the true redshift distribution and the redshift error distribution, and because of this, the plotted points are not independent.  For the majority of sources in the $z=1-7$ range, this error distribution function is $\delta z / (1+z)\approx0.28$.  The distribution peaks at a higher redshift than found by \cite{herve} and due to the favourable $k$-correction of observing at 850\,$\mu$m, this distribution may more accurately reflect the true redshift distribution of the \planck{} overdensities.  If most of the \planck{} overdensities are in fact physically associated structures, then those sources found by \cite{herve} are probably at somewhat higher redshifts and have warmer dust temperatures than assumed.  Conversely, we may be detecting colder components of these structures at 850\,$\mu$m.  However, the redshift distribution of the SCUBA-2-selected \planck{} overdensity sources is not significantly different than those within the CLS UDS field, other than a factor of roughly $4$ increase in the number of sources.  Unfortunately, this finding does not help to disentangle the fraction of \planck{} sources that are line-of-sight enhancements rather than being physically associated.

Using sources within the \planck{} beams (defined to be the area above half the peak flux density within each \planck{} map) and the CLS UDS, we can assess what fraction of the \planck{} 850\,$\mu$m flux density of the 51 fields we recover with SCUBA-2.  To do this, we must first quantify the expected flux density contribution for a blank field (we will subtract this from the total SCUBA-2 flux density of our sources found to be within the Planck beams).  Applying the same selection function of our survey to the CLS UDS source list, we recover a total flux density of 0.39\,Jy per 1.05\,deg$^2$ (the area of the Planck beams).  From the Cosmic Background Explorer ({\it COBE}) satellite data, the total flux density of this area should be 46\,Jy \cite[although admittedly with about a 30 per cent uncertainty,][]{cobe}, thus we only recover about 1 per cent of the far-IR background in the CLS UDS field.  \planck{} measured a total 850\,$\mu$m flux density of 23.6\,Jy within the 51 fields observed \citep[using CMB and Galactic cirrus cleaned maps,][]{planckselection}.  Summing the SCUBA-2 850\,$\mu$m flux densities of the sources found to be within the Planck beams, we recover a total flux density of 1.53\,Jy.  Subtracting off the expected blank-field contribution estimated from the CLS UDS field, we end up with a 1.14\,Jy enhancement above the background signal, and we conclude that we recover around 5 per cent of the \planck{} flux density within these fields.  Comparing the relative fractions of recovered flux densities shows that the 850\,$\mu$m number counts are enhanced by a larger amount at high flux densities compared to fainter sources for our survey.  One must also consider that the \planck{} flux densities may have a significant flux-boosting effect, due to their low S/N and the large area used to find these overdensities; thus the fraction we recover with SCUBA-2 may be underestimated.

\begin{figure*}
\begin{center}
\includegraphics[width=10cm,trim = 25mm 0mm 15mm 10mm, clip, angle=-90]{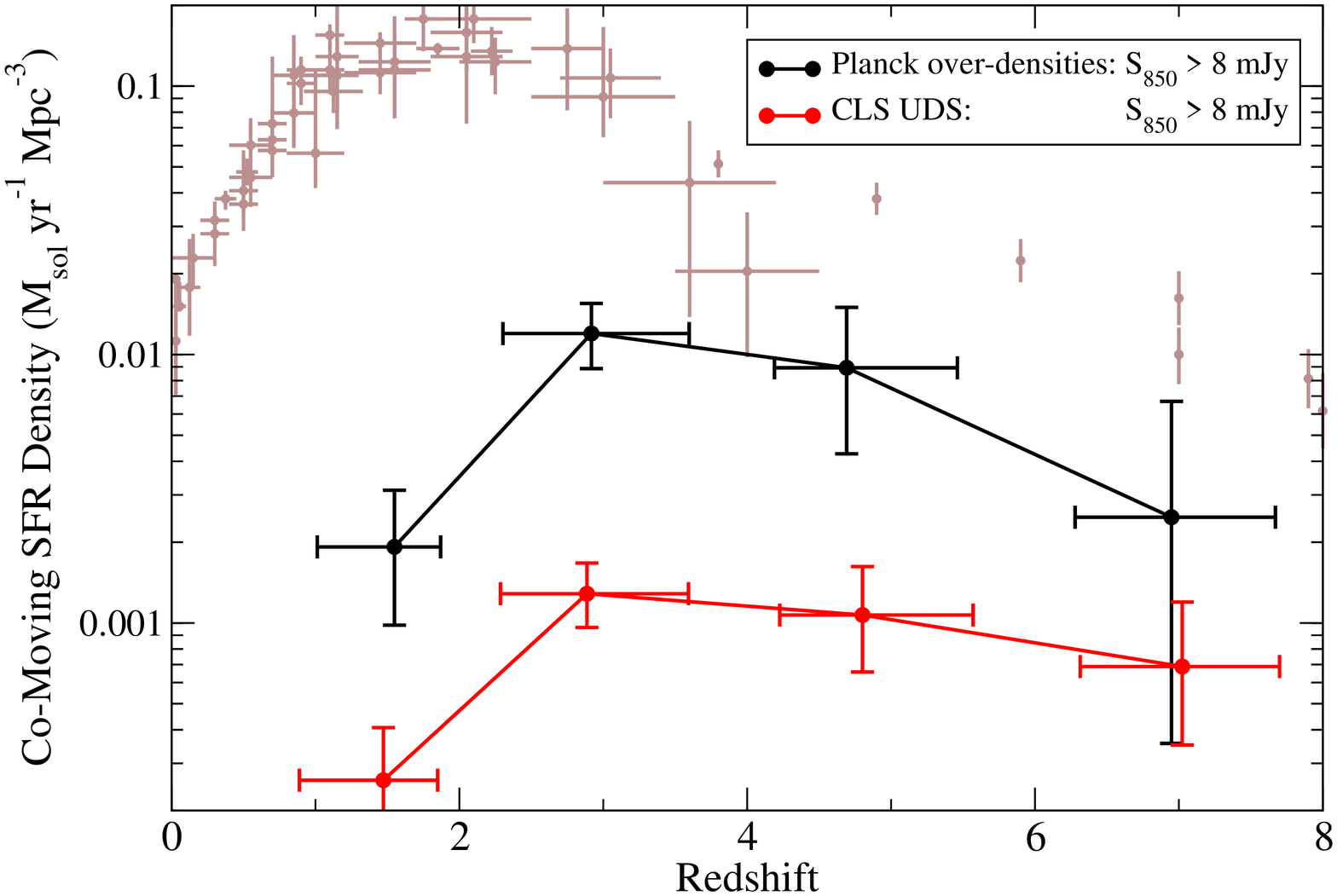}
\end{center}
\caption[Co-moving SFR density versus redshift for SCUBA-2-detected sources with flux densities greater than 8\,mJy, assuming a dust temperature prior of $34^{+13}_{-9}$\,K.]{Co-moving SFR density versus redshift for SCUBA-2-detected sources with flux densities greater than 8\,mJy assuming a dust temperature prior of $34^{+13}_{-9}$\,K.  This sample is only expected to have approximately 1 contributing spurious SCUBA-2 source.  For comparison, we also show results from fitting to sources within the CLS UDS field, using the same technique, with a similar selection and flux-boosting effects.  An order of magnitude increase in star-formation rate density is seen across all redshifts.  The light brown points are measurements of the global co-moving star-formation density from an assortment of sources, as compiled by \cite{madau}.}
\label{sfrvsredshift}
\end{figure*}

In Fig.~\ref{sfrvsredshift} we estimate the co-moving SFR density for SCUBA-2-detected \planck{} overdensity sources within the \planck{} beam and CLS UDS fields with flux densities greater than 8\,mJy (and flux density uncertainty $<2$\,mJy) assuming a dust temperature prior of $34^{+13}_{-9}$\,K and a conversion factor of $1.08\times 10^{-10} \mathrm{M}_\odot \mathrm{yr}^{-1} \mathrm{L}_\odot^{-1}$ for a Chabrier
IMF \citep[as in, e.g.][]{swinbank}.  We see up to an order of magnitude increase in the SFR density in the \planck{} overdensity fields in comparison with the CLS UDS field, across a broad range of redshifts.  Again, this plot is a convolution of the true co-moving SFR density with the redshift error function.  The plot is generated by bootstrapping the MCMC chains and is corrected for contributions from spurious sources by subtracting negative SCUBA-2 sources (although with our chosen flux cut, we only have 1 negative source to subtract from the \planck{} overdensity fields and no correction is applied to the CLS UDS field).  We add noise to the CLS SCUBA-2 flux densities in order to simulate the \planck{} overdensity selection function, similar to above, but here we match the flux density uncertainty distribution of regions with uncertainty below 2\,mJy.  With this more strict flux density cut-off, only 0.11\,deg$^2$ and 45 sources of the \planck{} overdensity fields remain.  In comparison, an average of 70 CLS UDS sources and the majority of the original survey area are still used.  This procedure gives us uncorrected number counts of 409 and 67 sources per square degree brighter than 8\,mJy for the \planck{} overdensity and CLS UDS fields, respectively, i.e. the \planck{} fields contain approximately 6 times higher surface density of 850\,$\mu$m sources than random parts of the sky.

\section{Conclusions}

We have followed up 61 \planck{} high-$z$ candidates using SCUBA-2 on the JCMT.  Of these, 10 are strong gravitational lenses, discussed in \cite{gems}.  The other 51 of the fields are referred to as ``\planck{} overdensities'' and are possible proto-cluster candidates.  We have first of all confirmed, using SCUBA-2 850\,$\mu$m data, that these fields do indeed represent regions of enhanced submm galaxy surface density on the sky.  We have used the {\sc SEDeblend} method from \cite{alesspaper} and the available SCUBA-2 and \herschel{}-SPIRE observations to constrain the redshifts and far-IR luminosities of 172 SCUBA-2-detected sources.  To do so, we have assumed a dust temperature prior of $34^{+13}_{-9}$\,K and find that we achieve a redshift uncertainty of $\delta z / (1+z)\approx0.28$ for the majority of sources.

We show that these \planck{}-selected fields have a factor of roughly 6 more sources above 8\,mJy at 850\,$\mu$m than blank field surveys, and that most of these sources are between a redshift of 2 and 4.  These sources appear to follow approximately the same redshift distribution as those found in blank field surveys.  We resolve around 5 per cent of the total \planck{} flux density.  Given the same selection function, blank field surveys only recover about 1 per cent of the extragalactic far-IR background, and thus we conclude that the number counts in these fields are more enhanced at high flux densities ($>8$\,mJy) than at lower flux densities.  We show that the SFR density in these fields are approximately an order of magnitude higher for sources $>8$\,mJy for redshifts out to $z\sim6$.  Determining if the structures are in fact physically associated, or if some are simply chance alignments, will require spectroscopic redshifts at either optical or submm wavelengths.  Such follow-up programmes are already underway.

\section*{Acknowledgements}

This research has been supported by the Natural Sciences and Engineering Research Council of Canada.
The James Clerk Maxwell Telescope is operated by the East Asian Observatory on behalf of The National Astronomical Observatory of Japan, Academia Sinica Institute of Astronomy and Astrophysics, the Korea Astronomy and Space Science Institute, the National Astronomical Observatories of China and the Chinese Academy of Sciences (Grant No. XDB09000000), with additional funding support from the Science and Technology Facilities Council of the United Kingdom and participating universities in the United Kingdom and Canada.
SPIRE has been developed by a consortium of institutes led by Cardiff University (UK) and including Univ. Lethbridge (Canada); NAOC (China); CEA, LAM (France); IFSI, Univ. Padua (Italy); IAC (Spain); Stockholm Observatory (Sweden); Imperial College London, RAL, UCL-MSSL, UKATC, Univ. Sussex (UK); and Caltech, JPL, NHSC, Univ. Colorado (USA). This development has been supported by national funding agencies: CSA (Canada); NAOC (China); CEA, CNES, CNRS (France); ASI (Italy); MCINN (Spain); SNSB (Sweden); STFC, UKSA (UK); and NASA (USA).
This research has made use of data from the HerMES project (\url{http://hermes.sussex.ac.uk/}), a Herschel Key Programme utilising Guaranteed Time from the SPIRE instrument team, ESAC scientists and a mission scientist. 
The HerMES data were accessed through the HeDaM database (\url{http://hedam.oamp.fr}) operated by CeSAM and hosted by the Laboratoire d'Astrophysique de Marseille.
This research used the facilities of the Canadian Astronomy Data Centre operated by the National Research Council of Canada with the support of the Canadian Space Agency. 
This research was enabled in part by support provided by WestGrid (\url{www.westgrid.ca}) and Compute Canada Calcul Canada (\url{www.computecanada.ca}).

\bibliography{references}

\begin{thebibliography}{}

\bibitem[\protect\citeauthoryear{{Blain} \& {Longair}}{{Blain} \&
  {Longair}}{1993}]{blain}
{Blain} A.~W.,  {Longair} M.~S.,  1993, \mnras, 264, 509

\bibitem[\protect\citeauthoryear{{Bleem} et~al.,}{{Bleem}
  et~al.}{2015}]{bleem2015}
{Bleem} L.~E.,  et~al., 2015, \apjs, 216, 27

\bibitem[\protect\citeauthoryear{{Braglia} et~al.,}{{Braglia}
  et~al.}{2011}]{braglia2011}
{Braglia} F.~G.,  et~al., 2011, \mnras, 412, 1187

\bibitem[\protect\citeauthoryear{{Braglia}, {Pierini}, {Biviano} \&
  {B{\"o}hringer}}{{Braglia} et~al.}{2009}]{braglia2009}
{Braglia} F.~G.,  {Pierini} D.,  {Biviano} A.,    {B{\"o}hringer} H.,  2009,
  \aap, 500, 947

\bibitem[\protect\citeauthoryear{{Canameras} et~al.,}{{Canameras}
  et~al.}{2015}]{gems}
{Canameras} R.,  et~al., 2015, ArXiv e-prints

\bibitem[\protect\citeauthoryear{{Chapin}, {Berry}, {Gibb}, {Jenness}, {Scott},
  {Tilanus}, {Economou} \& {Holland}}{{Chapin} et~al.}{2013}]{smurf}
{Chapin} E.~L.,  {Berry} D.~S.,  {Gibb} A.~G.,  {Jenness} T.,  {Scott} D.,
  {Tilanus} R.~P.~J.,  {Economou} F.,    {Holland} W.~S.,  2013, \mnras, 430,
  2545

\bibitem[\protect\citeauthoryear{{Chapman}, {Blain}, {Smail} \&
  {Ivison}}{{Chapman} et~al.}{2005}]{chapman2005}
{Chapman} S.~C.,  {Blain} A.~W.,  {Smail} I.,    {Ivison} R.~J.,  2005, \apj,
  622, 772

\bibitem[\protect\citeauthoryear{{Clements} et~al.,}{{Clements}
  et~al.}{2014}]{clements2014}
{Clements} D.~L.,  et~al., 2014, \mnras, 439, 1193

\bibitem[\protect\citeauthoryear{{Clements}, {Sutherland}, {McMahon} \&
  {Saunders}}{{Clements} et~al.}{1996}]{clements1996}
{Clements} D.~L.,  {Sutherland} W.~J.,  {McMahon} R.~G.,    {Saunders} W.,
  1996, \mnras, 279, 477

\bibitem[\protect\citeauthoryear{{Combes} et~al.,}{{Combes}
  et~al.}{2012}]{combes}
{Combes} F.,  et~al., 2012, \aap, 538, L4

\bibitem[\protect\citeauthoryear{{Coppin} et~al.,}{{Coppin}
  et~al.}{2006}]{coppin2006}
{Coppin} K.,  et~al., 2006, \mnras, 372, 1621

\bibitem[\protect\citeauthoryear{{Daddi}, {Dickinson}, {Morrison}, {Chary},
  {Cimatti}, {Elbaz}, {Frayer}, {Renzini}, {Pope}, {Alexander}, {Bauer},
  {Giavalisco}, {Huynh}, {Kurk} \& {Mignoli}}{{Daddi} et~al.}{2007}]{daddi2007}
{Daddi} E.,  {Dickinson} M.,  {Morrison} G.,  {Chary} R.,  {Cimatti} A.,
  {Elbaz} D.,  {Frayer} D.,  {Renzini} A.,  {Pope} A.,  {Alexander} D.~M.,
  {Bauer} F.~E.,  {Giavalisco} M.,  {Huynh} M.,  {Kurk} J.,    {Mignoli} M.,
  2007, \apj, 670, 156

\bibitem[\protect\citeauthoryear{{Dempsey} \& {Friberg}}{{Dempsey} \&
  {Friberg}}{2008}]{wvm}
{Dempsey} J.,  {Friberg} P.,  2008, SPIE, 7012, 137

\bibitem[\protect\citeauthoryear{{Dempsey} et~al.,}{{Dempsey}
  et~al.}{2013}]{scuba2calibration}
{Dempsey} J.~T.,  et~al., 2013, \mnras, 430, 2534

\bibitem[\protect\citeauthoryear{{Dempsey}, {Friberg}, {Jenness}, {Bintley} \&
  {Holland}}{{Dempsey} et~al.}{2010}]{dempsey2010}
{Dempsey} J.~T.,  {Friberg} P.,  {Jenness} T.,  {Bintley} D.,    {Holland}
  W.~S.,  2010 Vol.~7741 of Society of Photo-Optical Instrumentation Engineers
  (SPIE) Conference Series, {Extinction correction and on-sky calibration of
  SCUBA-2}.
p.~1

\bibitem[\protect\citeauthoryear{{Eisenhardt}, {Brodwin}, {Gonzalez},
  {Stanford}, {Stern}, {Barmby}, {Brown}, {Dawson}, {Dey}, {Doi}, {Galametz},
  {Jannuzi}, {Kochanek}, {Meyers}, {Morokuma} \& {Moustakas}}{{Eisenhardt}
  et~al.}{2008}]{iracsurvey}
{Eisenhardt} P.~R.~M.,  {Brodwin} M.,  {Gonzalez} A.~H.,  {Stanford} S.~A.,
  {Stern} D.,  {Barmby} P.,  {Brown} M.~J.~I.,  {Dawson} K.,  {Dey} A.,  {Doi}
  M.,  {Galametz} A.,  {Jannuzi} B.~T.,  {Kochanek} C.~S.,  {Meyers} J.,
  {Morokuma} T.,    {Moustakas} L.~A.,  2008, \apj, 684, 905

\bibitem[\protect\citeauthoryear{{Elbaz} et~al.,}{{Elbaz}
  et~al.}{2011}]{elbaz2011}
{Elbaz} D.,  et~al., 2011, \aap, 533, A119

\bibitem[\protect\citeauthoryear{{Fixsen}, {Dwek}, {Mather}, {Bennett} \&
  {Shafer}}{{Fixsen} et~al.}{1998}]{cobe}
{Fixsen} D.~J.,  {Dwek} E.,  {Mather} J.~C.,  {Bennett} C.~L.,    {Shafer}
  R.~A.,  1998, \apj, 508, 123

\bibitem[\protect\citeauthoryear{{Franceschini}, {Toffolatti}, {Mazzei},
  {Danese} \& {de Zotti}}{{Franceschini} et~al.}{1991}]{franceschini}
{Franceschini} A.,  {Toffolatti} L.,  {Mazzei} P.,  {Danese} L.,    {de Zotti}
  G.,  1991, \aaps, 89, 285

\bibitem[\protect\citeauthoryear{{Fu} et~al.,}{{Fu}  et~al.}{2012}]{fu}
{Fu} H.,  et~al., 2012, \apj, 753, 134

\bibitem[\protect\citeauthoryear{Gibb \& Jenness}{Gibb \&
  Jenness}{2010}]{oracdr}
Gibb A.,  Jenness T.,  2010, Processing SCUBA-2 Data with ORAC-DR, Starlink
  User Note 264, Version 1.0.0,
  \url{http://www.starlink.ac.uk/docs/sun264.htx/sun264.html}

\bibitem[\protect\citeauthoryear{{Griffin} et~al.,}{{Griffin}
  et~al.}{2010}]{spire}
{Griffin} M.~J.,  et~al., 2010, \aa, 518, L3

\bibitem[\protect\citeauthoryear{{Hasselfield} et~al.,}{{Hasselfield}
  et~al.}{2013}]{actsz}
{Hasselfield} M.,  et~al., 2013, \jcap, 7, 8

\bibitem[\protect\citeauthoryear{Hastings}{Hastings}{1970}]{mcmc2}
Hastings W.~K.,  1970, Biometrika, 57, 97

\bibitem[\protect\citeauthoryear{{Holland}, {Robson}, {Gear}, {Cunningham},
  {Lightfoot}, {Jenness}, {Ivison}, {Stevens}, {Ade}, {Griffin}, {Duncan},
  {Murphy} \& {Naylor}}{{Holland} et~al.}{1999}]{holland}
{Holland} W.~S.,  {Robson} E.~I.,  {Gear} W.~K.,  {Cunningham} C.~R.,
  {Lightfoot} J.~F.,  {Jenness} T.,  {Ivison} R.~J.,  {Stevens} J.~A.,  {Ade}
  P.~A.~R.,  {Griffin} M.~J.,  {Duncan} W.~D.,  {Murphy} J.~A.,    {Naylor}
  D.~A.,  1999, \mnras, 303, 659

\bibitem[\protect\citeauthoryear{{Lawrence} et~al.,}{{Lawrence}
  et~al.}{2007}]{uds}
{Lawrence} A.,  et~al., 2007, \mnras, 379, 1599

\bibitem[\protect\citeauthoryear{{MacKenzie}, {Scott} \&
  {Swinbank}}{{MacKenzie} et~al.}{2015}]{alesspaper}
{MacKenzie} T.,  {Scott} D.,    {Swinbank} M.,  2015, ArXiv e-prints

\bibitem[\protect\citeauthoryear{{MacKenzie} et~al.,}{{MacKenzie}
  et~al.}{2014}]{Mackenzie2014}
{MacKenzie} T.~P.,  et~al., 2014, \mnras, 445, 201

\bibitem[\protect\citeauthoryear{{Madau} \& {Dickinson}}{{Madau} \&
  {Dickinson}}{2014}]{madau}
{Madau} P.,  {Dickinson} M.,  2014, ARA\&A, 52, 415

\bibitem[\protect\citeauthoryear{{Metropolis}, {Rosenbluth}, {Rosenbluth},
  {Teller} \& {Teller}}{{Metropolis} et~al.}{1953}]{mcmc1}
{Metropolis} N.,  {Rosenbluth} A.~W.,  {Rosenbluth} M.~N.,  {Teller} A.~H.,
  {Teller} E.,  1953, JCP, 21, 1087

\bibitem[\protect\citeauthoryear{{Murphy} Jr., {Armus}, {Matthews}, {Soifer},
  {Mazzarella}, {Shupe}, {Strauss} \& {Neugebauer}}{{Murphy}
  et~al.}{1996}]{murphy1996}
{Murphy} Jr. T.~W.,  {Armus} L.,  {Matthews} K.,  {Soifer} B.~T.,  {Mazzarella}
  J.~M.,  {Shupe} D.~L.,  {Strauss} M.~A.,    {Neugebauer} G.,  1996, \aj, 111,
  1025

\bibitem[\protect\citeauthoryear{{Negrello} et~al.,}{{Negrello}
  et~al.}{2010}]{negrello}
{Negrello} M.,  et~al., 2010, Science, 330, 800

\bibitem[\protect\citeauthoryear{{Nguyen} et~al.,}{{Nguyen}
  et~al.}{2010}]{confusion}
{Nguyen} H.~T.,  et~al., 2010, \aap, 518, L5

\bibitem[\protect\citeauthoryear{{Noeske} et~al.,}{{Noeske}
  et~al.}{2007}]{noeske2007}
{Noeske} K.~G.,  et~al., 2007, \apjl, 660, L43

\bibitem[\protect\citeauthoryear{{Oliver} et~al.,}{{Oliver}
  et~al.}{2012}]{hermespaper}
{Oliver} S.~J.,  et~al., 2012, \mnras, 424, 1614

\bibitem[\protect\citeauthoryear{{Ott}}{{Ott}}{2010}]{hipe}
{Ott} S.,  2010, in {Mizumoto} Y.,  {Morita} K.-I.,   {Ohishi} M.,  eds,
  Astronomical Data Analysis Software and Systems XIX Vol.~434 of Astronomical
  Society of the Pacific Conference Series, {The Herschel Data Processing
  System -- HIPE and Pipelines -- Up and Running Since the Start of the
  Mission}.
p.~139

\bibitem[\protect\citeauthoryear{{Pascale} et~al.,}{{Pascale}
  et~al.}{2008}]{pascale2008}
{Pascale} E.,  et~al., 2008, \apj, 681, 400

\bibitem[\protect\citeauthoryear{{Pilbratt}, {Riedinger}, {Passvogel}, {Crone},
  {Doyle}, {Gageur}, {Heras}, {Jewell}, {Metcalfe}, {Ott} \&
  {Schmidt}}{{Pilbratt} et~al.}{2010}]{pilbratt2010}
{Pilbratt} G.~L.,  {Riedinger} J.~R.,  {Passvogel} T.,  {Crone} G.,  {Doyle}
  D.,  {Gageur} U.,  {Heras} A.~M.,  {Jewell} C.,  {Metcalfe} L.,  {Ott} S.,
  {Schmidt} M.,  2010, \aap, 518, L1

\bibitem[\protect\citeauthoryear{{Planck Collaboration VII}}{{Planck
  Collaboration VII}}{2011}]{ercsc}
{Planck Collaboration VII} 2011, \aap, 536, A7

\bibitem[\protect\citeauthoryear{{Planck Collaboration VIII}}{{Planck
  Collaboration VIII}}{2011}]{plancksz1}
{Planck Collaboration VIII} 2011, \aap, 536, A8

\bibitem[\protect\citeauthoryear{{Planck Collaboration XIII}}{{Planck
  Collaboration XIII}}{2015}]{planckcosmology}
{Planck Collaboration XIII} 2015, ArXiv e-prints

\bibitem[\protect\citeauthoryear{{Planck Collaboration XX}}{{Planck
  Collaboration XX}}{2014}]{plancksz3}
{Planck Collaboration XX} 2014, \aap, 571, A20

\bibitem[\protect\citeauthoryear{{Planck Collaboration XXIX}}{{Planck
  Collaboration XXIX}}{2014}]{plancksz2}
{Planck Collaboration XXIX} 2014, \aap, 571, A29

\bibitem[\protect\citeauthoryear{{Planck Collaboration XXVII}}{{Planck
  Collaboration XXVII}}{2015}]{herve}
{Planck Collaboration XXVII} 2015, ArXiv e-prints

\bibitem[\protect\citeauthoryear{{Planck Collaboration XXVIII}}{{Planck
  Collaboration XXVIII}}{2014}]{pccs}
{Planck Collaboration XXVIII} 2014, AAP, 571, A28

\bibitem[\protect\citeauthoryear{{Planck Collaboration XXXIX}}{{Planck
  Collaboration XXXIX}}{2015}]{planckselection}
{Planck Collaboration XXXIX} 2015, ArXiv e-prints

\bibitem[\protect\citeauthoryear{{Sanders} \& {Mirabel}}{{Sanders} \&
  {Mirabel}}{1996}]{sanders1996}
{Sanders} D.~B.,  {Mirabel} I.~F.,  1996, \araa, 34, 749

\bibitem[\protect\citeauthoryear{{Scott} et~al.,}{{Scott}
  et~al.}{2010}]{scott2010}
{Scott} K.~S.,  et~al., 2010, \mnras, 405, 2260

\bibitem[\protect\citeauthoryear{{Simpson} et~al.,}{{Simpson}
  et~al.}{2015}]{simpson2015}
{Simpson} J.,  et~al., 2015, ArXiv e-prints

\bibitem[\protect\citeauthoryear{{Simpson} et~al.,}{{Simpson}
  et~al.}{2014}]{simpson}
{Simpson} J.~M.,  et~al., 2014, \apj, 788, 125

\bibitem[\protect\citeauthoryear{{Swinbank} et~al.,}{{Swinbank}
  et~al.}{2014}]{swinbank}
{Swinbank} A.~M.,  et~al., 2014, \mnras, 438, 1267

\bibitem[\protect\citeauthoryear{{Verdugo}, {Ziegler} \& {Gerken}}{{Verdugo}
  et~al.}{2008}]{verdugo2008}
{Verdugo} M.,  {Ziegler} B.~L.,    {Gerken} B.,  2008, \aap, 486, 9

\end{thebibliography}

\onecolumn
\begingroup
\scriptsize
\setlength{\LTleft}{-20cm plus -1fill}
\setlength{\LTright}{\LTleft}
\begin{center}
\begin{longtable}{ccccccc} 
\caption[SCUBA-2-detected sources within the \planck{} proto-cluster candidate fields.]{SCUBA-2-detected sources within the \planck{} proto-cluster candidate fields.  Flux density uncertainties at 850\,$\mu$m are instrumental noise only and do not include confusion noise.  We report the median values from the MCMC chains and 68 per cent confidence intervals for both far-IR luminosities and photometric redshifts.  Photometric redshift fits for sources with redshifts greater than about 6.5 are not possible, due to the peak of the SED being located outside the wavelength coverage.  For these sources, we provide the 16th per centile of the distributions as a 1$\sigma$ lower confidence limit.}
\label{sourcelist}\\
\hline
Galaxy ID & RA & Dec & $S_{850}$ & $z_{\mathrm{phot}}$ & Far-IR luminosity & In \planck{} beam\\ 
 & J2000 & J2000 & (mJy) & & (L$_\odot$) & \\
\hline \\
\endfirsthead
\endhead
\hline
Gal ID & RA & Dec & S$_{850}$ & $z_{\mathrm{phot}}$ & Far-IR Luminosity & In Planck\\ 
 & J2000 & J2000 & (mJy) & & (L$_\odot$) & Beam\\
\hline \\
\endhead
PCCS\_G045.7-41.2-0 & 21:39:51.055 & $-$8:47:16.80 & $13.1\pm2.5$ & $4.1\substack{+1.6\\ -1.3}$ & $1.7\substack{+1.9\\ -1.1}\times10^{13}$  & Y  \\
PCCS\_G045.7-41.2-1 & 21:39:30.820 & $-$8:44:08.79 & $11.1\pm2.4$ & $2.7\substack{+1.0\\ -0.9}$ & $1.3\substack{+1.5\\ -0.8}\times10^{13}$  & Y  \\
PCCS\_G045.7-41.2-2 & 21:39:47.817 & $-$8:44:20.80 & $9.9\pm2.1$ & $0.1\substack{+0.1\\ -0.1}$ & $9.7\substack{+46.4\\ -8.4}\times10^{10}$  & Y  \\
PCCS\_G045.7-41.2-3 & 21:39:29.200 & $-$8:46:04.78 & $13.1\pm3.2$ & $3.9\substack{+1.4\\ -1.2}$ & $1.4\substack{+1.3\\ -0.8}\times10^{13}$  & N  \\
PCCS\_G059.1-67.1-0 & 23:26:25.977 & $-$15:28:05.40 & $14.5\pm1.6$ & $3.3\substack{+1.3\\ -1.1}$ & $1.5\substack{+1.6\\ -0.9}\times10^{13}$  & Y  \\
PCCS\_G059.1-67.1-1 & 23:26:01.346 & $-$15:30:45.32 & $18.1\pm3.8$ & $>7.0$ & $2.7\substack{+6.3\\ -1.2}\times10^{13}$  & N  \\
PCCS\_G059.1-67.1-2 & 23:26:41.749 & $-$15:28:57.36 & $13.3\pm2.8$ & $3.3\substack{+1.2\\ -1.0}$ & $1.2\substack{+1.2\\ -0.7}\times10^{13}$  & N  \\
PCCS\_G059.1-67.1-3 & 23:26:47.004 & $-$15:27:17.34 & $14.8\pm3.5$ & $>7.1$ & $2.2\substack{+5.2\\ -1.0}\times10^{13}$  & N  \\
PCCS\_G073.4-57.5-0 & 23:14:42.344 & $-$4:16:40.20 & $10.4\pm1.8$ & $2.6\substack{+1.0\\ -0.9}$ & $9.4\substack{+11.0\\ -6.1}\times10^{12}$  & Y  \\
PCCS\_G073.4-57.5-1 & 23:14:42.611 & $-$4:20:00.20 & $13.6\pm2.5$ & $4.8\substack{+1.7\\ -1.4}$ & $1.6\substack{+1.4\\ -0.8}\times10^{13}$  & N  \\
PCCS\_G073.4-57.5-2 & 23:14:41.809 & $-$4:17:44.20 & $8.3\pm2.0$ & $>3.6$ & $9.8\substack{+9.9\\ -5.5}\times10^{12}$  & Y  \\
PCCS\_G073.4-57.5-3 & 23:14:34.589 & $-$4:17:00.20 & $7.2\pm1.8$ & $3.7\substack{+1.5\\ -1.2}$ & $7.9\substack{+8.2\\ -4.6}\times10^{12}$  & Y  \\
PHZ\_G038.0-51.5-0 & 22:08:50.400 & $-$17:55:47.90 & $15.9\pm2.0$ & $2.7\substack{+1.1\\ -1.0}$ & $1.9\substack{+2.2\\ -1.3}\times10^{13}$  & Y  \\
PHZ\_G038.0-51.5-1 & 22:08:48.438 & $-$17:57:35.90 & $10.3\pm2.0$ & $3.1\substack{+1.2\\ -1.1}$ & $1.1\substack{+1.3\\ -0.7}\times10^{13}$  & Y  \\
PHZ\_G038.0-51.5-2 & 22:08:45.353 & $-$17:59:27.90 & $10.4\pm2.4$ & $2.3\substack{+0.9\\ -0.9}$ & $9.5\substack{+12.3\\ -6.5}\times10^{12}$  & Y  \\
PHZ\_G067.2-63.8-0 & 23:24:24.799 & $-$10:43:46.08 & $22.7\pm3.5$ & $3.1\substack{+1.2\\ -1.1}$ & $2.0\substack{+2.3\\ -1.3}\times10^{13}$  & N  \\
PHZ\_G067.2-63.8-10 & 23:24:27.248 & $-$10:51:10.07 & $11.8\pm2.8$ & $4.1\substack{+1.6\\ -1.3}$ & $1.3\substack{+1.2\\ -0.7}\times10^{13}$  & N  \\
PHZ\_G067.2-63.8-11 & 23:24:12.314 & $-$10:48:06.10 & $5.9\pm1.5$ & $>5.4$ & $7.9\substack{+8.2\\ -3.9}\times10^{12}$  & Y  \\
PHZ\_G067.2-63.8-1 & 23:24:12.043 & $-$10:46:14.10 & $11.1\pm1.8$ & $2.6\substack{+1.0\\ -0.9}$ & $1.2\substack{+1.4\\ -0.8}\times10^{13}$  & Y  \\
PHZ\_G067.2-63.8-2 & 23:24:23.718 & $-$10:50:18.08 & $12.6\pm2.1$ & $3.0\substack{+1.3\\ -1.0}$ & $1.6\substack{+2.1\\ -1.0}\times10^{13}$  & N  \\
PHZ\_G067.2-63.8-3 & 23:24:01.728 & $-$10:45:26.09 & $13.1\pm2.2$ & $4.9\substack{+1.8\\ -1.5}$ & $1.6\substack{+1.4\\ -0.9}\times10^{13}$  & Y  \\
PHZ\_G067.2-63.8-4 & 23:24:00.643 & $-$10:45:06.09 & $11.9\pm2.2$ & $2.7\substack{+1.3\\ -1.0}$ & $1.3\substack{+2.0\\ -0.9}\times10^{13}$  & Y  \\
PHZ\_G067.2-63.8-5 & 23:24:16.658 & $-$10:48:30.10 & $7.8\pm1.7$ & $3.6\substack{+1.5\\ -1.1}$ & $8.9\substack{+9.4\\ -5.1}\times10^{12}$  & Y  \\
PHZ\_G067.2-63.8-6 & 23:24:06.614 & $-$10:47:14.10 & $7.1\pm1.5$ & $2.3\substack{+0.9\\ -0.8}$ & $8.1\substack{+9.9\\ -5.5}\times10^{12}$  & Y  \\
PHZ\_G067.2-63.8-7 & 23:23:56.023 & $-$10:51:30.08 & $10.7\pm2.4$ & $1.6\substack{+0.8\\ -0.7}$ & $1.3\substack{+2.1\\ -1.0}\times10^{13}$  & N  \\
PHZ\_G067.2-63.8-8 & 23:24:01.453 & $-$10:52:06.09 & $8.9\pm2.1$ & $6.3\substack{+2.3\\ -2.1}$ & $1.1\substack{+0.9\\ -0.6}\times10^{13}$  & Y  \\
PHZ\_G067.2-63.8-9 & 23:23:59.014 & $-$10:45:34.08 & $8.6\pm2.1$ & $3.5\substack{+1.4\\ -1.1}$ & $8.7\substack{+8.7\\ -5.1}\times10^{12}$  & Y  \\
PHZ\_G103.1-73.6-0 & 0:28:48.011 & $-$11:35:52.39 & $9.0\pm2.2$ & $>4.0$ & $1.1\substack{+1.0\\ -0.6}\times10^{13}$  & Y  \\
PHZ\_G106.8-83.3-0 & 0:43:25.677 & $-$20:36:20.29 & $22.8\pm1.7$ & $3.6\substack{+1.3\\ -1.3}$ & $2.5\substack{+2.7\\ -1.6}\times10^{13}$  & Y  \\
PHZ\_G106.8-83.3-10 & 0:43:12.001 & $-$20:37:36.29 & $7.0\pm1.6$ & $3.6\substack{+1.5\\ -1.2}$ & $7.6\substack{+8.3\\ -4.4}\times10^{12}$  & Y  \\
PHZ\_G106.8-83.3-1 & 0:43:17.415 & $-$20:36:00.30 & $10.2\pm1.4$ & $2.2\substack{+0.9\\ -0.8}$ & $1.5\substack{+1.9\\ -1.0}\times10^{13}$  & Y  \\
PHZ\_G106.8-83.3-11 & 0:43:29.099 & $-$20:38:44.28 & $8.1\pm2.0$ & $4.9\substack{+2.0\\ -1.6}$ & $9.7\substack{+8.5\\ -5.2}\times10^{12}$  & Y  \\
PHZ\_G106.8-83.3-12 & 0:43:17.700 & $-$20:36:48.30 & $5.9\pm1.4$ & $2.3\substack{+1.0\\ -0.8}$ & $7.9\substack{+10.9\\ -5.1}\times10^{12}$  & Y  \\
PHZ\_G106.8-83.3-2 & 0:43:14.281 & $-$20:36:12.30 & $10.1\pm1.4$ & $2.4\substack{+1.1\\ -0.9}$ & $9.5\substack{+13.6\\ -6.4}\times10^{12}$  & Y  \\
PHZ\_G106.8-83.3-3 & 0:43:21.689 & $-$20:36:40.30 & $9.1\pm1.5$ & $5.2\substack{+2.3\\ -1.7}$ & $1.1\substack{+0.9\\ -0.6}\times10^{13}$  & Y  \\
PHZ\_G106.8-83.3-4 & 0:43:02.320 & $-$20:34:04.26 & $12.4\pm2.1$ & $2.6\substack{+1.0\\ -0.9}$ & $1.5\substack{+1.7\\ -1.0}\times10^{13}$  & Y  \\
PHZ\_G106.8-83.3-5 & 0:43:32.230 & $-$20:36:48.26 & $12.0\pm2.1$ & $2.7\substack{+1.1\\ -0.8}$ & $1.5\substack{+1.7\\ -0.9}\times10^{13}$  & Y  \\
PHZ\_G106.8-83.3-6 & 0:43:41.064 & $-$20:37:20.20 & $19.1\pm3.9$ & $>7.0$ & $3.2\substack{+9.9\\ -1.5}\times10^{13}$  & N  \\
PHZ\_G106.8-83.3-7 & 0:43:04.598 & $-$20:34:24.27 & $9.4\pm2.0$ & $3.3\substack{+1.2\\ -1.1}$ & $1.1\substack{+1.1\\ -0.7}\times10^{13}$  & Y  \\
PHZ\_G106.8-83.3-8 & 0:43:04.882 & $-$20:34:52.27 & $8.7\pm1.9$ & $4.7\substack{+2.1\\ -1.6}$ & $1.0\substack{+0.8\\ -0.5}\times10^{13}$  & Y  \\
PHZ\_G106.8-83.3-9 & 0:43:32.807 & $-$20:40:44.26 & $13.9\pm3.1$ & $3.9\substack{+1.6\\ -1.2}$ & $1.3\substack{+1.2\\ -0.7}\times10^{13}$  & Y  \\
PHZ\_G119.4-76.6-0 & 0:48:10.840 & $-$13:45:56.30 & $24.5\pm2.2$ & $3.1\substack{+1.1\\ -1.1}$ & $2.9\substack{+3.1\\ -1.9}\times10^{13}$  & Y  \\
PHZ\_G119.4-76.6-1 & 0:47:52.723 & $-$13:41:56.28 & $10.7\pm1.9$ & $2.6\substack{+1.0\\ -1.1}$ & $1.9\substack{+2.1\\ -1.4}\times10^{13}$  & Y  \\
PHZ\_G119.4-76.6-2 & 0:48:10.016 & $-$13:44:28.30 & $10.8\pm2.4$ & $3.7\substack{+1.5\\ -1.2}$ & $1.2\substack{+1.3\\ -0.7}\times10^{13}$  & Y  \\
PHZ\_G119.4-76.6-3 & 0:47:59.311 & $-$13:40:32.30 & $7.4\pm1.7$ & $>5.5$ & $1.1\substack{+1.6\\ -0.5}\times10^{13}$  & Y  \\
PHZ\_G119.4-76.6-4 & 0:48:02.604 & $-$13:40:12.30 & $7.0\pm1.7$ & $>2.7$ & $8.6\substack{+11.3\\ -5.2}\times10^{12}$  & Y  \\
PHZ\_G119.4-76.6-5 & 0:47:53.273 & $-$13:41:12.28 & $7.6\pm1.8$ & $>5.5$ & $1.1\substack{+1.2\\ -0.5}\times10^{13}$  & Y  \\
PHZ\_G119.4-76.6-6 & 0:48:08.093 & $-$13:37:24.30 & $9.2\pm2.2$ & $>5.6$ & $1.2\substack{+1.1\\ -0.6}\times10^{13}$  & N  \\
PHZ\_G119.4-76.6-7 & 0:47:45.864 & $-$13:39:36.26 & $10.0\pm2.5$ & $2.3\substack{+0.9\\ -0.8}$ & $9.6\substack{+12.1\\ -6.4}\times10^{12}$  & N  \\
PHZ\_G119.4-76.6-8 & 0:48:13.584 & $-$13:43:12.29 & $8.1\pm2.0$ & $>6.9$ & $1.1\substack{+1.2\\ -0.5}\times10^{13}$  & Y  \\
PHZ\_G132.6-81.1-0 & 0:57:48.895 & $-$18:19:23.50 & $13.3\pm2.1$ & $3.3\substack{+1.4\\ -1.2}$ & $1.5\substack{+1.8\\ -1.0}\times10^{13}$  & Y  \\
PHZ\_G132.6-81.1-1 & 0:57:52.265 & $-$18:19:03.49 & $8.6\pm2.1$ & $>4.8$ & $1.1\substack{+1.0\\ -0.6}\times10^{13}$  & Y  \\
PHZ\_G171.1-78.7-0 & 1:27:01.926 & $-$19:19:41.60 & $14.6\pm2.1$ & $3.1\substack{+1.1\\ -1.1}$ & $1.6\substack{+1.7\\ -1.0}\times10^{13}$  & Y  \\
PHZ\_G171.1-78.7-1 & 1:26:50.339 & $-$19:20:13.56 & $15.4\pm3.6$ & $5.4\substack{+2.4\\ -1.7}$ & $1.7\substack{+1.5\\ -0.9}\times10^{13}$  & Y  \\
PHZ\_G171.1-78.7-2 & 1:27:08.708 & $-$19:18:57.60 & $9.0\pm2.1$ & $4.9\substack{+2.4\\ -1.7}$ & $1.1\substack{+1.0\\ -0.6}\times10^{13}$  & Y  \\
PHZ\_G171.1-78.7-3 & 1:27:08.708 & $-$19:19:01.60 & $8.7\pm2.1$ & $4.0\substack{+2.0\\ -1.3}$ & $1.1\substack{+1.0\\ -0.6}\times10^{13}$  & Y  \\
PHZ\_G173.9+57.0-0 & 10:28:38.124 & 43:25:37.69 & $8.5\pm1.9$ & $3.7\substack{+1.6\\ -1.2}$ & $9.5\substack{+9.9\\ -5.6}\times10^{12}$  & Y  \\
PHZ\_G173.9+57.0-1 & 10:28:48.771 & 43:24:53.70 & $8.7\pm2.0$ & $2.4\substack{+1.0\\ -0.9}$ & $1.1\substack{+1.5\\ -0.8}\times10^{13}$  & Y  \\
PHZ\_G176.6+59.0-0 & 10:36:56.556 & 41:27:22.40 & $15.8\pm2.6$ & $3.3\substack{+1.2\\ -1.2}$ & $1.9\substack{+1.9\\ -1.2}\times10^{13}$  & Y  \\
PHZ\_G176.6+59.0-1 & 10:37:05.451 & 41:27:30.38 & $12.5\pm3.0$ & $2.1\substack{+1.0\\ -0.8}$ & $1.4\substack{+2.2\\ -1.0}\times10^{13}$  & Y  \\
PHZ\_G214.1+48.3-0 & 9:52:34.268 & 19:08:18.69 & $15.7\pm3.1$ & $6.4\substack{+2.2\\ -2.0}$ & $1.8\substack{+1.4\\ -0.9}\times10^{13}$  & Y  \\
PHZ\_G214.1+48.3-1 & 9:52:39.349 & 19:08:30.67 & $14.9\pm3.2$ & $>6.9$ & $2.2\substack{+5.1\\ -1.0}\times10^{13}$  & Y  \\
PHZ\_G214.1+48.3-2 & 9:52:09.714 & 19:06:26.66 & $15.2\pm3.6$ & $5.6\substack{+2.1\\ -1.6}$ & $1.7\substack{+1.3\\ -0.8}\times10^{13}$  & N  \\
Planck18p194-0 & 8:30:46.455 & 19:36:47.19 & $19.6\pm1.9$ & $3.3\substack{+1.4\\ -1.1}$ & $2.2\substack{+2.7\\ -1.4}\times10^{13}$  & Y  \\
Planck18p194-1 & 8:30:54.382 & 19:37:31.20 & $14.9\pm1.7$ & $3.3\substack{+1.4\\ -1.2}$ & $1.8\substack{+2.2\\ -1.1}\times10^{13}$  & Y  \\
Planck18p194-2 & 8:30:51.551 & 19:37:55.20 & $10.7\pm1.6$ & $5.3\substack{+2.1\\ -1.7}$ & $1.3\substack{+1.1\\ -0.7}\times10^{13}$  & Y  \\
Planck18p194-3 & 8:30:41.073 & 19:39:43.18 & $13.7\pm2.5$ & $4.9\substack{+1.7\\ -1.5}$ & $1.6\substack{+1.5\\ -0.9}\times10^{13}$  & Y  \\
Planck18p194-4 & 8:31:04.287 & 19:34:23.18 & $13.7\pm3.1$ & $3.4\substack{+1.3\\ -1.1}$ & $1.5\substack{+1.5\\ -0.9}\times10^{13}$  & N  \\
Planck18p194-5 & 8:30:40.228 & 19:36:15.17 & $8.5\pm2.0$ & $>6.7$ & $1.3\substack{+2.3\\ -0.6}\times10^{13}$  & Y  \\
Planck18p194-6 & 8:30:51.268 & 19:37:31.20 & $6.8\pm1.7$ & $3.2\substack{+1.2\\ -1.0}$ & $8.9\substack{+9.7\\ -5.4}\times10^{12}$  & Y  \\
Planck18p194-7 & 8:30:48.719 & 19:37:55.20 & $6.5\pm1.6$ & $4.0\substack{+1.5\\ -1.2}$ & $8.1\substack{+7.8\\ -4.6}\times10^{12}$  & Y  \\
Planck18p735-0 & 1:58:48.085 & $-$7:52:43.50 & $8.5\pm2.0$ & $2.8\substack{+1.2\\ -1.0}$ & $1.3\substack{+1.7\\ -0.9}\times10^{13}$  & Y  \\
Planck24p194-0 & 8:40:40.588 & 22:12:37.60 & $8.3\pm1.6$ & $2.8\substack{+1.1\\ -0.9}$ & $1.2\substack{+1.4\\ -0.8}\times10^{13}$  & Y  \\
PLCK\_G006.1+61.8-0 & 14:33:47.1 & 12:12:60.00 & $16.0\pm2.8$ & $2.8\substack{+1.1\\ -1.0}$ & $1.3\substack{+1.6\\ -0.9}\times10^{13}$  & Y  \\
PLCK\_G006.1+61.8-1 & 14:33:39.817 & 12:14:52.00 & $14.0\pm3.0$ & $3.7\substack{+1.4\\ -1.1}$ & $1.7\substack{+1.8\\ -1.0}\times10^{13}$  & Y  \\
PLCK\_G009.8+72.6-0 & 13:59:19.151 & 19:19:15.97 & $18.7\pm2.2$ & $3.2\substack{+1.2\\ -1.1}$ & $2.2\substack{+2.3\\ -1.4}\times10^{13}$  & Y  \\
PLCK\_G009.8+72.6-1 & 13:59:02.479 & 19:19:32.00 & $10.1\pm2.2$ & $5.3\substack{+2.0\\ -1.6}$ & $1.2\substack{+1.0\\ -0.6}\times10^{13}$  & Y  \\
PLCK\_G009.8+72.6-2 & 13:59:28.188 & 19:16:27.92 & $11.9\pm2.9$ & $3.9\substack{+1.4\\ -1.2}$ & $1.3\substack{+1.2\\ -0.7}\times10^{13}$  & N  \\
PLCK\_G009.8+72.6-3 & 13:58:57.958 & 19:18:15.99 & $11.1\pm2.7$ & $5.8\substack{+3.0\\ -2.8}$ & $1.2\substack{+1.1\\ -0.7}\times10^{13}$  & Y  \\
PLCK\_G049.6-42.9-0 & 21:51:38.531 & $-$7:05:06.90 & $9.9\pm1.6$ & $2.4\substack{+1.0\\ -0.9}$ & $1.2\substack{+1.6\\ -0.8}\times10^{13}$  & Y  \\
PLCK\_G049.6-42.9-1 & 21:51:41.219 & $-$7:05:54.90 & $7.8\pm1.9$ & $2.7\substack{+1.1\\ -0.9}$ & $7.4\substack{+8.4\\ -4.4}\times10^{12}$  & Y  \\
PLCK\_G049.6-42.9-2 & 21:51:33.694 & $-$7:05:02.90 & $6.5\pm1.6$ & $>5.6$ & $8.7\substack{+8.5\\ -4.3}\times10^{12}$  & Y  \\
PLCK\_G056.7+62.6-0 & 14:54:39.298 & 34:43:28.00 & $15.9\pm2.7$ & $2.9\substack{+1.2\\ -1.0}$ & $1.8\substack{+2.1\\ -1.2}\times10^{13}$  & Y  \\
PLCK\_G056.7+62.6-1 & 14:54:38.649 & 34:46:24.00 & $10.7\pm2.4$ & $3.2\substack{+1.3\\ -1.1}$ & $1.3\substack{+1.6\\ -0.8}\times10^{13}$  & Y  \\
PLCK\_G056.7+62.6-2 & 14:54:28.259 & 34:47:11.98 & $14.4\pm3.3$ & $3.6\substack{+1.4\\ -1.1}$ & $1.2\substack{+1.2\\ -0.7}\times10^{13}$  & Y  \\
PLCK\_G068.3+31.9-0 & 17:33:13.960 & 42:42:21.70 & $18.8\pm2.8$ & $2.2\substack{+0.9\\ -0.8}$ & $2.5\substack{+3.1\\ -1.6}\times10^{13}$  & Y  \\
PLCK\_G068.3+31.9-1 & 17:33:32.479 & 42:45:09.63 & $14.4\pm3.6$ & $3.3\substack{+1.2\\ -1.0}$ & $1.3\substack{+1.3\\ -0.7}\times10^{13}$  & N  \\
PLCK\_G075.1+33.2-0 & 17:29:51.000 & 48:31:35.00 & $13.1\pm2.7$ & $6.2\substack{+1.6\\ -2.2}$ & $1.5\substack{+1.0\\ -0.8}\times10^{13}$  & Y  \\
PLCK\_G077.7+32.6-0 & 17:33:47.863 & 50:44:56.17 & $14.9\pm3.7$ & $2.0\substack{+0.8\\ -0.7}$ & $8.4\substack{+10.5\\ -5.5}\times10^{12}$  & N  \\
PLCK\_G078.9+48.2-0 & 15:56:11.488 & 50:04:32.77 & $12.8\pm2.4$ & $4.1\substack{+1.5\\ -1.4}$ & $1.4\substack{+1.3\\ -0.8}\times10^{13}$  & Y  \\
PLCK\_G078.9+48.2-1 & 15:55:36.170 & 50:04:28.63 & $14.2\pm3.3$ & $>5.4$ & $1.9\substack{+3.2\\ -1.0}\times10^{13}$  & Y  \\
PLCK\_G082.5+38.4-0 & 16:55:59.511 & 54:30:00.89 & $18.1\pm2.0$ & $>7.0$ & $3.3\substack{+15.3\\ -1.5}\times10^{13}$  & Y  \\
PLCK\_G082.5+38.4-1 & 16:55:31.952 & 54:30:36.77 & $11.6\pm2.8$ & $2.0\substack{+0.9\\ -0.7}$ & $7.7\substack{+9.8\\ -5.0}\times10^{12}$  & Y  \\
PLCK\_G082.5+38.4-2 & 16:55:39.750 & 54:32:28.85 & $10.1\pm2.5$ & $2.5\substack{+0.9\\ -0.8}$ & $8.2\substack{+8.9\\ -5.0}\times10^{12}$  & N  \\
PLCK\_G083.3+51.0-0 & 15:33:13.312 & 51:47:39.00 & $12.2\pm2.2$ & $3.4\substack{+1.3\\ -1.1}$ & $1.3\substack{+1.4\\ -0.8}\times10^{13}$  & Y  \\
PLCK\_G083.3+51.0-1 & 15:32:51.293 & 51:52:06.92 & $16.5\pm3.5$ & $4.1\substack{+1.5\\ -1.3}$ & $1.7\substack{+1.6\\ -1.0}\times10^{13}$  & Y  \\
PLCK\_G083.3+51.0-2 & 15:32:57.793 & 51:46:54.97 & $13.0\pm2.9$ & $3.2\substack{+1.7\\ -1.0}$ & $9.2\substack{+10.0\\ -5.4}\times10^{12}$  & Y  \\
PLCK\_G084.0-71.5-0 & 0:04:16.645 & $-$12:18:09.58 & $16.5\pm2.6$ & $4.2\substack{+1.5\\ -1.3}$ & $1.8\substack{+1.8\\ -1.0}\times10^{13}$  & Y  \\
PLCK\_G084.0-71.5-1 & 0:04:17.194 & $-$12:14:05.58 & $15.5\pm3.6$ & $>4.8$ & $1.9\substack{+1.7\\ -1.0}\times10^{13}$  & N  \\
PLCK\_G084.0-71.5-2 & 0:04:30.019 & $-$12:17:53.60 & $9.1\pm2.2$ & $>5.6$ & $1.2\substack{+1.0\\ -0.6}\times10^{13}$  & Y  \\
PLCK\_G091.9+43.0-0 & 16:09:59.845 & 60:19:52.00 & $17.2\pm3.2$ & $3.6\substack{+1.3\\ -1.2}$ & $1.5\substack{+1.6\\ -0.9}\times10^{13}$  & Y  \\
PLCK\_G091.9+43.0-1 & 16:10:14.926 & 60:19:15.96 & $15.5\pm3.1$ & $3.4\substack{+1.3\\ -1.1}$ & $1.8\substack{+1.9\\ -1.1}\times10^{13}$  & Y  \\
PLCK\_G093.6+55.9-0 & 14:44:05.173 & 54:16:45.00 & $17.3\pm3.7$ & $>5.7$ & $2.4\substack{+2.5\\ -1.2}\times10^{13}$  & N  \\
PLCK\_G093.6+55.9-1 & 14:43:56.475 & 54:21:16.97 & $11.5\pm2.5$ & $3.1\substack{+1.3\\ -1.0}$ & $1.6\substack{+1.9\\ -1.0}\times10^{13}$  & Y  \\
PLCK\_G132.9-76.0-0 & 1:01:00.549 & $-$13:17:54.25 & $16.1\pm3.8$ & $2.6\substack{+1.0\\ -0.9}$ & $1.6\substack{+1.9\\ -1.1}\times10^{13}$  & N  \\
PLCK\_G144.1+81.0-0 & 12:35:34.282 & 35:28:40.50 & $13.0\pm2.7$ & $3.4\substack{+1.3\\ -1.0}$ & $1.3\substack{+1.4\\ -0.7}\times10^{13}$  & Y  \\
PLCK\_G144.1+81.0-1 & 12:35:46.730 & 35:30:08.45 & $14.2\pm3.5$ & $>6.3$ & $2.0\substack{+2.4\\ -1.0}\times10^{13}$  & N  \\
PLCK\_G160.7+41.0-0 & 9:07:54.534 & 56:03:10.89 & $22.3\pm3.9$ & $3.9\substack{+1.5\\ -1.3}$ & $2.9\substack{+3.2\\ -1.7}\times10^{13}$  & Y  \\
PLCK\_G162.1-59.3-0 & 2:06:51.367 & $-$2:16:05.80 & $8.1\pm1.6$ & $2.5\substack{+1.0\\ -0.9}$ & $9.1\substack{+10.7\\ -5.9}\times10^{12}$  & Y  \\
PLCK\_G162.1-59.3-1 & 2:06:39.892 & $-$2:11:21.80 & $14.3\pm3.2$ & $3.1\substack{+1.1\\ -1.1}$ & $2.0\substack{+2.1\\ -1.3}\times10^{13}$  & Y  \\
PLCK\_G162.1-59.3-2 & 2:06:39.090 & $-$2:16:57.80 & $8.4\pm2.0$ & $3.1\substack{+1.2\\ -1.1}$ & $9.8\substack{+10.9\\ -6.1}\times10^{12}$  & N  \\
PLCK\_G162.1-59.3-3 & 2:06:49.232 & $-$2:15:49.80 & $6.8\pm1.6$ & $>4.5$ & $8.9\substack{+7.9\\ -4.5}\times10^{12}$  & Y  \\
PLCK\_G165.8+45.3-0 & 9:30:34.209 & 51:28:06.19 & $14.2\pm3.5$ & $>5.7$ & $1.9\substack{+1.7\\ -0.9}\times10^{13}$  & N  \\
PLCK\_G173.8+59.3-0 & 10:40:31.859 & 42:43:23.00 & $12.6\pm2.1$ & $3.2\substack{+1.3\\ -1.1}$ & $1.2\substack{+1.4\\ -0.8}\times10^{13}$  & Y  \\
PLCK\_G173.8+59.3-1 & 10:40:30.765 & 42:48:11.00 & $17.1\pm3.4$ & $>7.3$ & $2.9\substack{+13.7\\ -1.3}\times10^{13}$  & Y  \\
PLCK\_G177.0+35.9-0 & 8:30:58.465 & 43:40:11.16 & $11.0\pm2.3$ & $5.6\substack{+1.9\\ -1.6}$ & $1.4\substack{+1.1\\ -0.7}\times10^{13}$  & N  \\
PLCK\_G177.0+35.9-1 & 8:31:13.948 & 43:38:03.20 & $8.1\pm1.7$ & $1.7\substack{+0.7\\ -0.7}$ & $6.0\substack{+8.2\\ -4.3}\times10^{12}$  & Y  \\
PLCK\_G177.0+35.9-2 & 8:31:18.741 & 43:39:39.18 & $8.6\pm1.8$ & $5.5\substack{+2.8\\ -2.0}$ & $1.0\substack{+0.9\\ -0.5}\times10^{13}$  & Y  \\
PLCK\_G177.0+35.9-3 & 8:31:41.579 & 43:37:50.95 & $14.9\pm3.4$ & $>5.3$ & $1.9\substack{+1.6\\ -0.9}\times10^{13}$  & N  \\
PLCK\_G177.0+35.9-4 & 8:31:02.153 & 43:39:59.18 & $9.0\pm2.2$ & $0.5\substack{+0.5\\ -0.3}$ & $2.5\substack{+10.3\\ -2.3}\times10^{12}$  & Y  \\
PLCK\_G179.3+50.7-0 & 9:51:38.990 & 41:39:15.59 & $12.2\pm1.6$ & $2.8\substack{+1.2\\ -1.1}$ & $1.2\substack{+1.6\\ -0.8}\times10^{13}$  & Y  \\
PLCK\_G179.3+50.7-1 & 9:51:41.131 & 41:39:43.60 & $8.4\pm1.5$ & $2.9\substack{+1.2\\ -1.0}$ & $9.1\substack{+11.6\\ -5.6}\times10^{12}$  & Y  \\
PLCK\_G179.3+50.7-2 & 9:51:44.700 & 41:40:07.60 & $7.0\pm1.5$ & $3.7\substack{+1.7\\ -1.2}$ & $7.5\substack{+7.5\\ -4.2}\times10^{12}$  & Y  \\
PLCK\_G179.3+50.7-3 & 9:52:00.771 & 41:41:47.53 & $9.0\pm2.0$ & $4.3\substack{+1.6\\ -1.4}$ & $1.0\substack{+1.0\\ -0.6}\times10^{13}$  & Y  \\
PLCK\_G179.3+50.7-4 & 9:51:45.413 & 41:35:39.60 & $10.6\pm2.4$ & $>7.0$ & $1.6\substack{+3.5\\ -0.7}\times10^{13}$  & N  \\
PLCK\_G179.3+50.7-5 & 9:51:45.414 & 41:37:59.60 & $7.3\pm1.7$ & $3.2\substack{+1.4\\ -1.1}$ & $7.5\substack{+8.0\\ -4.5}\times10^{12}$  & Y  \\
PLCK\_G186.3-72.7-0 & 1:56:33.074 & $-$18:27:31.60 & $11.3\pm1.8$ & $3.0\substack{+1.3\\ -1.0}$ & $1.3\substack{+1.7\\ -0.8}\times10^{13}$  & Y  \\
PLCK\_G186.3-72.7-1 & 1:56:33.636 & $-$18:28:47.60 & $8.7\pm2.0$ & $5.2\substack{+2.4\\ -1.7}$ & $1.2\substack{+1.0\\ -0.6}\times10^{13}$  & Y  \\
PLCK\_G186.3-72.7-2 & 1:56:34.199 & $-$18:28:39.60 & $8.7\pm2.0$ & $2.9\substack{+1.2\\ -1.0}$ & $1.4\substack{+1.8\\ -0.9}\times10^{13}$  & Y  \\
PLCK\_G186.6+66.7-0 & 11:08:36.022 & 35:06:04.00 & $12.7\pm2.4$ & $4.8\substack{+1.8\\ -1.5}$ & $1.5\substack{+1.3\\ -0.8}\times10^{13}$  & Y  \\
PLCK\_G188.6-68.9-0 & 2:11:48.227 & $-$17:00:57.40 & $13.4\pm1.9$ & $2.4\substack{+1.1\\ -1.0}$ & $1.3\substack{+2.0\\ -1.0}\times10^{13}$  & Y  \\
PLCK\_G188.6-68.9-10 & 2:11:45.159 & $-$17:00:41.40 & $9.1\pm2.1$ & $2.5\substack{+1.1\\ -1.0}$ & $1.1\substack{+1.6\\ -0.8}\times10^{13}$  & Y  \\
PLCK\_G188.6-68.9-11 & 2:11:52.688 & $-$16:59:01.40 & $9.9\pm2.4$ & $4.2\substack{+2.2\\ -1.4}$ & $9.5\substack{+9.1\\ -5.3}\times10^{12}$  & Y  \\
PLCK\_G188.6-68.9-1 & 2:11:49.063 & $-$17:02:49.40 & $8.8\pm1.5$ & $2.5\substack{+1.0\\ -0.8}$ & $8.7\substack{+10.7\\ -5.3}\times10^{12}$  & Y  \\
PLCK\_G188.6-68.9-2 & 2:11:52.131 & $-$17:02:45.40 & $8.3\pm1.5$ & $>5.1$ & $1.2\substack{+1.2\\ -0.6}\times10^{13}$  & Y  \\
PLCK\_G188.6-68.9-3 & 2:11:38.745 & $-$17:01:09.38 & $10.4\pm2.0$ & $2.5\substack{+1.1\\ -0.9}$ & $1.2\substack{+1.6\\ -0.8}\times10^{13}$  & Y  \\
PLCK\_G188.6-68.9-4 & 2:11:33.720 & $-$17:04:21.36 & $11.6\pm2.4$ & $3.1\substack{+1.2\\ -1.1}$ & $1.4\substack{+1.5\\ -0.9}\times10^{13}$  & Y  \\
PLCK\_G188.6-68.9-5 & 2:11:44.878 & $-$17:05:57.40 & $9.2\pm2.0$ & $2.6\substack{+1.0\\ -0.9}$ & $9.9\substack{+11.7\\ -6.6}\times10^{12}$  & N  \\
PLCK\_G188.6-68.9-6 & 2:11:51.853 & $-$17:05:49.40 & $11.2\pm2.5$ & $3.0\substack{+1.2\\ -1.0}$ & $1.7\substack{+2.1\\ -1.1}\times10^{13}$  & N  \\
PLCK\_G188.6-68.9-7 & 2:11:42.649 & $-$17:00:57.39 & $8.9\pm2.0$ & $>5.7$ & $1.2\substack{+1.1\\ -0.6}\times10^{13}$  & Y  \\
PLCK\_G188.6-68.9-8 & 2:11:47.669 & $-$17:01:45.40 & $7.1\pm1.6$ & $3.1\substack{+1.1\\ -1.0}$ & $1.0\substack{+1.1\\ -0.6}\times10^{13}$  & Y  \\
PLCK\_G188.6-68.9-9 & 2:11:56.315 & $-$17:03:21.39 & $7.8\pm1.8$ & $4.6\substack{+2.2\\ -1.4}$ & $8.8\substack{+8.0\\ -4.7}\times10^{12}$  & Y  \\
PLCK\_G191.3+62.0-0 & 10:44:57.144 & 33:51:38.09 & $12.2\pm3.0$ & $1.6\substack{+0.8\\ -0.7}$ & $1.2\substack{+1.9\\ -0.9}\times10^{13}$  & Y  \\
PLCK\_G191.3+62.0-1 & 10:45:07.096 & 33:50:50.03 & $16.0\pm3.9$ & $3.3\substack{+1.3\\ -1.0}$ & $1.1\substack{+1.1\\ -0.6}\times10^{13}$  & Y  \\
PLCK\_G191.8-83.4-0 & 1:18:28.190 & $-$24:34:10.19 & $9.6\pm1.6$ & $4.1\substack{+1.6\\ -1.4}$ & $1.0\substack{+1.0\\ -0.6}\times10^{13}$  & Y  \\
PLCK\_G191.8-83.4-1 & 1:18:23.791 & $-$24:34:26.17 & $12.8\pm2.2$ & $3.4\substack{+1.3\\ -1.1}$ & $1.4\substack{+1.6\\ -0.8}\times10^{13}$  & Y  \\
PLCK\_G191.8-83.4-2 & 1:18:21.148 & $-$24:36:42.15 & $10.9\pm1.9$ & $3.2\substack{+1.3\\ -1.1}$ & $1.1\substack{+1.2\\ -0.7}\times10^{13}$  & Y  \\
PLCK\_G191.8-83.4-3 & 1:18:30.241 & $-$24:35:38.19 & $9.1\pm1.6$ & $3.0\substack{+1.2\\ -1.0}$ & $1.2\substack{+1.5\\ -0.7}\times10^{13}$  & Y  \\
PLCK\_G191.8-83.4-4 & 1:18:25.253 & $-$24:37:30.17 & $9.4\pm1.8$ & $2.0\substack{+0.9\\ -0.8}$ & $1.2\substack{+1.8\\ -0.9}\times10^{13}$  & Y  \\
PLCK\_G191.8-83.4-5 & 1:18:39.920 & $-$24:36:10.20 & $8.7\pm1.7$ & $2.1\substack{+0.9\\ -0.8}$ & $9.2\substack{+11.6\\ -6.4}\times10^{12}$  & Y  \\
PLCK\_G191.8-83.4-6 & 1:18:27.605 & $-$24:32:46.18 & $8.7\pm1.9$ & $>6.2$ & $1.3\substack{+1.3\\ -0.6}\times10^{13}$  & Y  \\
PLCK\_G191.8-83.4-7 & 1:18:36.693 & $-$24:34:38.20 & $6.8\pm1.5$ & $2.5\substack{+1.8\\ -1.0}$ & $7.7\substack{+10.2\\ -5.0}\times10^{12}$  & Y  \\
PLCK\_G191.8-83.4-8 & 1:18:41.680 & $-$24:36:46.19 & $7.5\pm1.8$ & $3.0\substack{+1.1\\ -1.0}$ & $8.7\substack{+9.5\\ -5.3}\times10^{12}$  & Y  \\
PLCK\_G191.8-83.4-9 & 1:18:36.986 & $-$24:34:42.20 & $6.0\pm1.5$ & $3.5\substack{+3.3\\ -1.6}$ & $6.8\substack{+7.2\\ -4.0}\times10^{12}$  & Y  \\
PLCK\_G201.1+50.7-0 & 9:53:11.581 & 27:54:30.40 & $9.2\pm1.7$ & $3.3\substack{+1.2\\ -1.1}$ & $1.1\substack{+1.2\\ -0.7}\times10^{13}$  & Y  \\
PLCK\_G201.1+50.7-1 & 9:53:08.864 & 27:55:38.39 & $8.3\pm1.8$ & $1.9\substack{+1.4\\ -0.8}$ & $9.2\substack{+12.5\\ -6.3}\times10^{12}$  & N  \\
PLCK\_G201.1+50.7-2 & 9:53:14.598 & 27:56:02.40 & $7.2\pm1.6$ & $3.7\substack{+1.5\\ -1.2}$ & $8.5\substack{+9.0\\ -5.0}\times10^{12}$  & Y  \\
PLCK\_G201.1+50.7-3 & 9:53:08.562 & 27:55:34.39 & $7.6\pm1.8$ & $3.0\substack{+2.5\\ -1.4}$ & $8.3\substack{+9.5\\ -5.4}\times10^{12}$  & N  \\
PLCK\_G213.0+65.9-0 & 11:04:38.213 & 24:36:35.50 & $12.3\pm3.0$ & $3.6\substack{+1.3\\ -1.1}$ & $1.3\substack{+1.2\\ -0.7}\times10^{13}$  & Y  \\
PLCK\_G213.0+65.9-1 & 11:04:44.075 & 24:33:39.48 & $13.7\pm3.4$ & $4.8\substack{+1.6\\ -1.3}$ & $1.5\substack{+1.2\\ -0.8}\times10^{13}$  & Y  \\
PLCK\_G223.9+41.2-0 & 9:37:14.190 & 10:00:05.39 & $16.2\pm1.9$ & $2.5\substack{+1.1\\ -0.9}$ & $1.6\substack{+2.2\\ -1.1}\times10^{13}$  & Y  \\
PLCK\_G223.9+41.2-1 & 9:36:45.214 & 10:01:45.37 & $14.0\pm2.3$ & $3.2\substack{+1.2\\ -1.1}$ & $1.4\substack{+1.5\\ -0.9}\times10^{13}$  & N  \\
PLCK\_G223.9+41.2-2 & 9:37:03.088 & 9:58:25.40 & $7.2\pm1.2$ & $1.2\substack{+0.7\\ -0.6}$ & $6.1\substack{+12.5\\ -4.7}\times10^{12}$  & Y  \\
PLCK\_G223.9+41.2-3 & 9:36:52.257 & 9:58:45.39 & $9.4\pm1.7$ & $2.9\substack{+1.2\\ -1.0}$ & $1.2\substack{+1.4\\ -0.7}\times10^{13}$  & Y  \\
PLCK\_G223.9+41.2-4 & 9:37:18.523 & 10:00:41.38 & $12.4\pm2.4$ & $>5.6$ & $1.7\substack{+1.3\\ -0.8}\times10^{13}$  & Y  \\
PLCK\_G223.9+41.2-5 & 9:37:01.192 & 9:58:29.40 & $5.9\pm1.2$ & $2.4\substack{+0.9\\ -0.9}$ & $6.3\substack{+7.6\\ -4.2}\times10^{12}$  & Y  \\
PLCK\_G223.9+41.2-6 & 9:36:42.512 & 9:56:21.36 & $10.2\pm2.3$ & $4.4\substack{+2.0\\ -1.4}$ & $1.1\substack{+0.9\\ -0.6}\times10^{13}$  & N  \\
PLCK\_G223.9+41.2-7 & 9:36:50.633 & 9:58:25.38 & $7.5\pm1.8$ & $4.3\substack{+1.9\\ -1.4}$ & $8.4\substack{+7.9\\ -4.7}\times10^{12}$  & Y  \\
PLCK\_G328.9+71.4-0 & 13:24:12.114 & 10:15:42.39 & $15.5\pm3.0$ & $3.4\substack{+1.3\\ -1.2}$ & $1.8\substack{+2.0\\ -1.1}\times10^{13}$  & N  \\
PLCK\_G328.9+71.4-1 & 13:24:03.171 & 10:12:22.40 & $10.5\pm2.1$ & $2.2\substack{+0.9\\ -0.8}$ & $1.3\substack{+1.7\\ -0.9}\times10^{13}$  & Y  \\
PLCK\_G328.9+71.4-2 & 13:23:44.744 & 10:14:10.37 & $13.6\pm3.1$ & $>6.8$ & $2.0\substack{+3.1\\ -0.9}\times10^{13}$  & Y  \\
PLCK\_G328.9+71.4-3 & 13:23:46.372 & 10:12:22.37 & $12.9\pm3.1$ & $>6.9$ & $2.0\substack{+4.6\\ -0.9}\times10^{13}$  & Y  \\
\hline
\end{longtable}
\end{center}
\twocolumn

\end{document}